\pdfoutput=1
\documentclass{article}

\usepackage{custom}
\usepackage{microtype}
\usepackage{graphicx}
\usepackage{subcaption}
\usepackage{booktabs} 

\usepackage{hyperref}

\usepackage[accepted]{icml2021}

\icmltitlerunning{Evaluating the Implicit Midpoint Integrator for Riemannian Manifold Hamiltonian Monte Carlo}

\begin{document}

\twocolumn[
\icmltitle{Evaluating the Implicit Midpoint Integrator for Riemannian Manifold Hamiltonian Monte Carlo}




\begin{icmlauthorlist}
\icmlauthor{James A. Brofos}{yale}
\icmlauthor{Roy R. Lederman}{yale}
\icmlcorrespondingauthor{James A. Brofos}{james.brofos@yale.edu}
\end{icmlauthorlist}

\icmlaffiliation{yale}{Department of Statistics and Data Science, Yale University}

\icmlkeywords{Machine Learning, ICML}

\vskip 0.3in
]



\printAffiliationsAndNotice{}  

\begin{abstract}
Riemannian manifold Hamiltonian Monte Carlo is traditionally carried out using the generalized leapfrog integrator. However, this integrator is not the only choice and other integrators yielding valid Markov chain transition operators may be considered. In this work, we examine the implicit midpoint integrator as an alternative to the generalized leapfrog integrator. We discuss advantages and disadvantages of the implicit midpoint integrator for Hamiltonian Monte Carlo, its theoretical properties, and an empirical assessment of the critical attributes of such an integrator for Hamiltonian Monte Carlo: energy conservation, volume preservation, and reversibility. Empirically, we find that while leapfrog iterations are faster, the implicit midpoint integrator has better energy conservation, leading to higher acceptance rates, as well as better conservation of volume and better reversibility, arguably yielding a more accurate sampling procedure.
\end{abstract}

\begin{figure}[t!]
    \centering
    \includegraphics[width=0.45\textwidth]{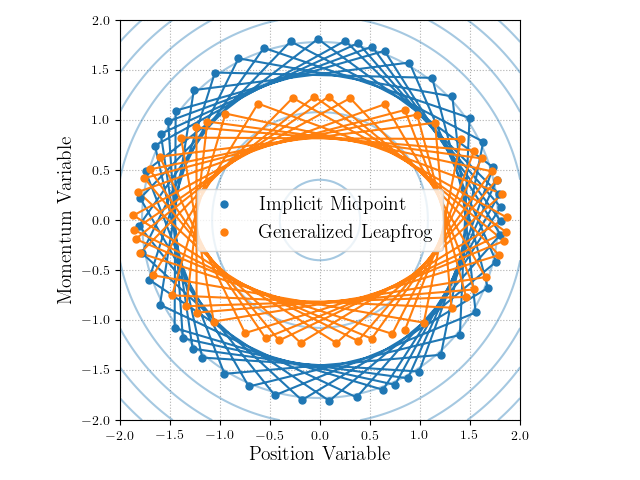}
    \caption{Trajectories computed by the leapfrog integrator and the implicit midpoint integrator for a quadratic, separable Hamiltonian $H(q, p) = q^\top q / 2 + p^\top p / 2$. Both integrators are stable but the leapfrog integrator deviates from the level sets of the Hamiltonian (faint blue circles) whereas every iterate of the implicit midpoint integrator lies on the same energy level set as the previous one.}
    \label{fig:gaussian-stability-energy-conservation}
\end{figure}

\section{Introduction}\label{sec:introduction}

Riemannian manifold Hamiltonian Monte Carlo (RMHMC) is a powerful algorithm for sampling from Bayesian posterior distributions \citep{doi:10.1111/j.1467-9868.2010.00765.x}. Given a log-posterior function $\mathcal{L} : \R^m\to\R$ and a Riemannian metric $\mathbb{G} : \R^m \to\R^{m\times m}$ (with the condition that $\mathbb{G}(q)$ is positive definite for each $q\in\R^m$), RMHMC considers the Hamiltonian dynamics corresponding to the Hamiltonian,
\begin{align}
    \label{eq:hamiltonian-energy} H(q, p) = -\mathcal{L}(q) + \frac{p^\top\mathbb{G}^{-1}(q)p}{2}  + \frac{\log \mathrm{det}(\mathbb{G}(q))}{2} .
\end{align}
Riemannian metrics are incorporated into HMC in order to precondition dynamics and more efficiently explore the distribution.
Irrespective of the choice of metric, the form of the Hamiltonian in \cref{eq:hamiltonian-energy} corresponds to a Gibbs distribution proportional to $\exp(-H(q, p))$ and tractable conditional distribution $p\vert q\sim \mathrm{Normal}(0, \mathbb{G}(q))$. 
However, the form of this Hamiltonian is such that it {\it cannot} be written as the sum of two functions, each a function of $q$ or $p$ alone; such a Hamiltonian is called ``non-separable.'' The presence of a non-separable Hamiltonian presents unique challenges for numerical integration.

The leapfrog integrator and its variants are a ubiquitous choice for the numerical integration of Hamiltonian mechanics for HMC; for instance see \cite{pmlr-v22-brubaker12,pmlr-v70-tripuraneni17a,1206.1901,Byrne_2013,softabs,doi:10.1111/j.1467-9868.2010.00765.x}, among many others. It may, therefore, not be apparent that numerical integrators other than the leapfrog method are applicable to HMC, provided that they exhibit two properties:
\begin{enumerate}[(i)]
    \itemsep0em
    \item The integrator has a unit Jacobian determinant so that it preserves volume in $(q, p)$-space.
    \item The integrator is symmetric under negation of the momentum variable.
\end{enumerate}

These properties are sufficient to prove that HMC satisfies detailed balance, which in turn establishes that the stationary distribution has density proportional to $\exp(-H(q, p))$; see \citet{1206.1901,10.5555/1162264}, or \cref{app:detailed-balance} for a proof.

Why does the choice of numerical integrator matter? There are at least three reasons.
\begin{enumerate}[(a)]
    \itemsep0em
    \item Numerical integrators differ with respect to energy conservation and stability. The acceptance probability of HMC depends on the energy conservation and the ability of the HMC proposal to use large integration steps depends on stability.
    \item Numerical integrators may only satisfy properties (i) and (ii) above {\it approximately}, particularly if the integrators are defined as solutions to implicitly-defined equations. These approximate solvers of implicit equations will be discussed in more details below. Typically, the error of these methods will depend on a convergence tolerance $\delta$ used to find fixed-points of the integration step (\cref{alg:fixed-point-solver}). For a non-zero convergence tolerance, the degree to which properties (i) and (ii) are violated will depend on the integrator and the tolerance.
    \item Numerical integrators will differ in their efficiency in the sense that there may be structural properties of the Hamiltonian system that the integrator exploits. More efficient integrators will exhibit higher effective sample sizes {\it per second} when used in HMC.
\end{enumerate}

The contribution of this work is to compare and contrast the generalized leapfrog integrator with the implicit midpoint method in application to RMHMC.
We consider RMHMC because the non-separable Hamiltonian necessitates elaborate integration schemes which require solving implicitly-defined equations; this is in contrast to Euclidean HMC with constant $\mathbb{G}$ which produces a separable Hamiltonian that can be integrated explicitly.
First, we compare the two integrators on the energy conservation, volume preservation, and reversibility as discussed in reason (a).  Second, we study the breakdown of {\it exact} satisfaction of properties (i) and (ii) in implicitly-defined integrators as described in reason (b). Third, we consider multiple variants of the  generalized leapfrog and implicit midpoint integrators that exhibit different efficiencies, relevant to reason (c). 
We conclude that the implicit midpoint integrator exhibits superior energy conservation, conservation of volume, and symmetry compared to the generalized leapfrog integrator. We explore inference in sophisticated Bayesian inference tasks wherein the implicit midpoint integrator is competitive with, or exceeds, the time-normalized performance of the generalized leapfrog method. We therefore argue that the implicit midpoint integrator is a procedure worth consideration in RMHMC.

\section{Background}

For Bayesian inference tasks, $\mathcal{L}$ is the sum of the log-likelihood and the log-prior; in this circumstance, the typical form of the Riemannian metric $\mathbb{G}$ is the sum of the Fisher information of the log-likelihood and negative Hessian of the log-prior; this choice of Riemannian metric is motivated by information geometry \citep{10.5555/3019383}. The Hamiltonian in \cref{eq:hamiltonian-energy} leads to the equations of motion,
\begin{align}
    \begin{split}
        \label{eq:position-evolution} \dot{q}_i &= \sum_{j=1}^m \mathbb{G}_{ij}^{-1}(q) p_j
    \end{split} \\
    \begin{split}
    \label{eq:momentum-evolution} \dot{p}_i &=  -\frac{\partial}{\partial q_i}\mathcal{L}(q) - \frac{1}{2} \mathrm{trace}\paren{\mathbb{G}^{-1}(q) \frac{\partial}{\partial q_i} \mathbb{G}(q)} \\
    & \qquad +~ \frac{1}{2} p^\top\mathbb{G}^{-1}(q) \frac{\partial}{\partial q_i}\mathbb{G}(q) \mathbb{G}^{-1}(q) p
    \end{split}
\end{align}

As stated in \cref{sec:introduction}, the standard integrator for RMHMC is the (generalized) leapfrog integrator. A naive implementation of a single step of the generalized leapfrog integrator with step-size $\epsilon$ and initial position $(q,p)$ is presented in \cref{alg:generalized-leapfrog}. Notice that \cref{eq:glf-momentum-i,eq:glf-position} are {\it implicitly defined} in the sense that the quantities appearing on the left-hand side also appear on the right-hand side; these equations are typically solved to a given tolerance $\delta\geq 0$ (in the sense defined in the fixed point iteration algorithm \cref{alg:fixed-point-solver}). When $\delta=0$, the generalized leapfrog integrator satisfies properties (i) and (ii), however, in practice, the tolerance is often chosen to be  larger than machine precision in order to reduce the number of fixed point iterations; therefore properties (i) and (ii) are no longer satisfied accurately.

The implicit midpoint method, an alternative to the generalize leapfrog integrator, is presented in \cref{alg:implicit-midpoint}; the implicit midpoint integrator also involves the solution to an implicitly-defined \cref{eq:implicit-midpoint}. When $\delta=0$, it is well-known that the implicit midpoint integrator satisfies property (i); see \citet{leimkuhler_reich_2005}. It also satisfies property (ii) for Hamiltonians of the form \cref{eq:hamiltonian-energy}; see \cref{app:momentum-negation-implicit-midpoint}.

\begin{algorithm}[t!]
\caption{{\bf (Fixed Point Iteration)} Procedure for solving the equation $z=f(z)$ via fixed point interation to a given tolerance.}
\label{alg:fixed-point-solver}
\begin{algorithmic}[1]
\STATE \textbf{Input}: Function $f :\R^m\to\R^m$, initial guess $z\in\R^m$, fixed point convergence tolerance $\delta \geq 0$.
\STATE Set $\Delta z = \infty$ and $z' = z$.
\STATE \textbf{While}: $\Delta z > \delta$ compute
\begin{align}
    z'' &= f(z') \\
    \Delta z &= \max_{i\in\set{1,\ldots, m}} \abs{z''_i - z'_i} \\
    z' &= z''
\end{align}
\STATE \textbf{Return}: $z'\in\R^m$.
\end{algorithmic}
\end{algorithm}
\begin{algorithm}[t!]
\caption{{\bf (G.L.F.(a))} The procedure for a single step of integrating Hamiltonian dynamics using the generalized leapfrog integrator.}
\label{alg:generalized-leapfrog}
\begin{algorithmic}[1]
\STATE \textbf{Input}: Hamiltonian $H:\R^m\times\R^m\to \R$, initial position and momentum variables $(q,p)\in\R^m\times\R^m$, integration step-size size $\epsilon\in\R$, fixed-point convergence tolerance $\delta \geq 0$.
\STATE Use \cref{alg:fixed-point-solver} with tolerance $\delta$ and initial guess $p$ to solve for $\bar{p}$,
\begin{align}
    \label{eq:glf-momentum-i} \bar{p} &\defeq \underbrace{p -\frac{\epsilon}{2} \nabla_q H(q, \bar{p})}_{f(\bar{p})}
\end{align}
\STATE Use \cref{alg:fixed-point-solver} with tolerance $\delta$ and initial guess $q$ to solve for $q'$,
\begin{align}
     \label{eq:glf-position} q' &\defeq \underbrace{q + \frac{\epsilon}{2} \paren{\nabla_p H(q, \bar{p}) + \nabla_p H(q', \bar{p})}}_{f(q')}
\end{align}
\STATE Compute the explicit update
\begin{align}
    \label{eq:glf-momentum-ii} p' &\defeq \bar{p} -\frac{\epsilon}{2} \nabla_qH(q', \bar{p})
\end{align}
\STATE \textbf{Return}: $(q', p')\in\R^m\times\R^m$.
\end{algorithmic}
\end{algorithm}
\begin{algorithm}[t!]
\caption{{\bf (I.M.(a))} The procedure for a single step of integrating Hamiltonian dynamics using the implicit midpoint integrator.}
\label{alg:implicit-midpoint}
\begin{algorithmic}[1]
\STATE \textbf{Input}: Hamiltonian $H:\R^m\times\R^m\to \R$, initial position and momentum variables $(q,p)\in\R^m\times\R^m$, integration step-size size $\epsilon\in\R$, fixed-point convergence tolerance $\delta \geq 0$.
\STATE Use \cref{alg:fixed-point-solver} with tolerance $\delta$ and initial guess $(q, p)$ to solve for $(q', p')$
\begin{align}
    \label{eq:implicit-midpoint} \begin{pmatrix} q' \\ p' \end{pmatrix} \defeq \underbrace{\begin{pmatrix} q \\ p \end{pmatrix} + \epsilon \begin{pmatrix} \nabla_p H(\bar{q}, \bar{p}) \\ -\nabla_q H(\bar{q},\bar{p}) \end{pmatrix}}_{f(q', p')}
\end{align}
where $\bar{q} \defeq (q' + q) / 2$ and $\bar{p} \defeq (p'+p)/2$.
\STATE \textbf{Return}: $(q', p')\in\R^m\times\R^m$.
\end{algorithmic}
\end{algorithm}

We turn now to discussing a theoretical property of numerical integrators related to conserved quantities.
\begin{definition}\label{def:conserved-quantity}
Let $z=(q, p)$ where the time evolution of $q_i$ is given by \cref{eq:position-evolution} and of $p_i$ by \cref{eq:momentum-evolution} for $i=1,\ldots, m$. A conserved quantity of $z$ is a real-valued function $z\mapsto \mathfrak{G}(z)$ for which $\frac{\mathrm{d}}{\mathrm{d}t} \mathfrak{G}(z) = 0$.
\end{definition}
It is important to notice that \cref{def:conserved-quantity} is a statement about the underlying dynamics and {\it has nothing to do with the integrator used to approximate these dynamics}. The properties of the integrators will be discussed in the next paragraph. 
For Hamiltonian systems, the canonical example of a conserved quantity is the Hamiltonian energy itself; see \citet{10.5555/1965128}. Hamiltonian flows are also  symplectic \citep{Hairer:1250576} which implies conservation of volume (in the same sense as that of property (i)).

A numerical integrator cannot preserve all of the conserved quantities as the underlying ODE, but it may be able to conserve some simple ones. The following two results may be found in \citet{leimkuhler_reich_2005}.
\begin{theorem}\label{thm:first-integral-generalized-leapfrog}
Let $z=(q, p)$. The generalized leapfrog integrator (\cref{alg:generalized-leapfrog}) with $\delta=0$ preserves any conserved quantity of the form $\mathfrak{G}(z) = q^\top\mathbf{A}p + \mathbf{b}^\top z$ where $\mathbf{A}\in\R^{m\times m}$ is a symmetric matrix and $\mathbf{b}\in\R^{2m}$.
\end{theorem}
\begin{theorem}\label{thm:first-integral-implicit-midpoint}
Let $z=(q, p)$. The implicit midpoint integrator (\cref{alg:implicit-midpoint}) with $\delta=0$ preserves any conserved quantity of the form $\mathfrak{G}(z) = z^\top\mathbf{A}z + \mathbf{b}^\top z$ where $\mathbf{A}\in\R^{2m\times 2m}$ is a symmetric matrix and $\mathbf{b}\in\R^{2m}$.
\end{theorem}
Notice that \cref{thm:first-integral-implicit-midpoint} contains a strictly more general class of conserved quantity than \cref{thm:first-integral-generalized-leapfrog}. 
We come now to a hypothesis that would justify the consideration of the implicit midpoint integrator within the context of HMC. Before stating the hypothesis, we provide some initial motivation for how the implicit midpoint integrator performs in the presence of a quadratic Hamiltonian. 
\begin{proposition}\label{prop:implicit-midpoint-quadratic-hamiltonian}
Let $H(q, p)\equiv H(z) = z^\top \mathbf{A}z$ be a quadratic Hamiltonian. Then, {\it for any step-size}, the proposals generated by Hamiltonian Monte Carlo using the implicit midpoint integrator with $\delta=0$ will be accepted.
\end{proposition}
A proof is given in \cref{app:implicit-midpoint-quadratic-hamiltonian}. Note, however, that perfect conservation of the Hamiltonian energy does {\it not} imply that the implicit midpoint integrator is the exact solution of the Hamilton's equations of motion. Nevertheless,
\Cref{prop:implicit-midpoint-quadratic-hamiltonian} suggests an important difference between the generalized leapfrog integrator and the implicit midpoint method in terms of their conservation properties. Although Bayesian posterior distributions are unlikely to be Gaussian, it is widely accepted that Gaussian approximations are useful. Such notions materialize, for example, in the central limit theorem and the Laplace approximation. We therefore speculate that the Gaussian case may be useful for providing intuition for the more general case we examine in our experimental results. Provided the posterior is approximately Gaussian, therefore, this leads us to the following hypothesis.

{\bf Hypothesis.} The implicit midpoint algorithm will exhibit higher acceptance probabilities than the generalized leapfrog integrator for the same step-size.

If true, and if the fixed point iterations required by the implicit midpoint procedure are not too burdensome relative to the generalized leapfrog integrator, then the higher acceptance rate may produce more favorable effective sample sizes for the Markov chain whose transitions are computed using the implicit midpoint algorithm. In this scenario, the implicit midpoint integrator may be worth consideration as an alternative to the generalized leapfrog integrator.

We wish to emphasize that the behavior of the implicit midpoint integrator in the presence of a quadratic Hamiltonian is not a definitive explanation of all differences in sampling behaviors that may arise when using it as a transition operator in RMHMC. However, we believe that the setting of quadratic Hamiltonians, corresponding to Gaussian densities, can provide helpful intuition. In \cref{sec:experimental-results} we will turn to the empirical evaluation of the implicit midpoint method to examine the extent to which this alternative integrator offers an advantage over the generalized leapfrog method in the non-Gaussian regime.

For a brief introduction to the stability of numerical integrators, see \cref{app:numerical-stability}.

\section{Related Work}\label{sec:related-work}

Most relevant to our discussion is \citet{pourzanjani2019implicit}. In this work, the authors examine the relationship between the (non-generalized) leapfrog integrator and the implicit midpoint integrator; the authors make the argument that the implicit midpoint integrator is more stable in the presence of posteriors whose dimensions exhibit large differences in their variability (``multi-scale''). The presence of multi-scale posterior dimensions necessitates a small step-size for the leapfrog integrator, which is found to be unnecessary for the implicit midpoint algorithm. As the authors note, however, ``RMHMC uses local Hessian evaluations of the potential energy surface to adaptively change this step-size based on the local curvature;'' therefore, their experiments instead focus on the circumstance where a constant mass matrix is utilized, corresponding to Euclidean HMC with no local adaptation of the step-size. Indeed, as observed in \citet{JMLR:v21:17-678}, the Fisher information captures the second-order geometry of the posterior and actually exhibits properties that make it preferable to the Hessian of the posterior in optimization. Therefore, the present work differs from \citet{pourzanjani2019implicit} in its focus on Riemannian geometry wherein the metric compensates (at least locally) for multi-scale dimensions; moreover, our {\it empirical} analysis of reversibility and volume preservation is, to the best of our knowledge, novel. Before proceeding to the experimental results, we note that stability alone {\it cannot} account for the high acceptance rate enjoyed by the implicit midpoint integrator: even in the regime wherein the generalized leapfrog integrator is stable, it is not able to perfectly conserve the Hamiltonian energy as the implicit midpoint integrator does. This phenomenon is visualized in \cref{fig:gaussian-stability-energy-conservation}.

\section{Experimental Results}\label{sec:experimental-results}

We turn now to evaluating the implicit midpoint integrator in several Bayesian inference tasks. We consider inference in a banana-shaped posterior, sampling from Neal's funnel distribution, a stochastic volatility model, and Bayesian inference in the Fitzhugh-Nagumo differential equation model. We have additional experimental results in our appendices. In \cref{app:experiment-quadratic-hamiltonian}, we seek to verify \cref{thm:first-integral-implicit-midpoint} in the presence of a truly quadratic Hamiltonian. In \cref{app:experiment-bayesian-logistic-regression}, we examine Bayesian inference in a logistic regression posterior. To define a stopping condition for the fixed point iterations used by the implicit midpoint and generalized leapfrog methods, we demand that the change in each coordinate be less than a threshold; we let  $\delta\in\set{1\times 10^{-9}, 1\times 10^{-6}, 1\times10^{-3}}$ when considering reversibility and volume preservation. When reporting performance metrics such as effective sample size, we report results corresponding to a threshold of $\delta=1\times 10^{-6}$. We implemented all methods in 64-bit precision using NumPy and SciPy \cite{harris2020array,2020SciPy-NMeth}. We compute effective sample sizes (ESS) using \citet{arviz_2019}. Additional experiments with a randomized number of integration steps are included in \cref{app:randomized-step-experimental-design}. Code for our experiments can be found at \url{https://github.com/JamesBrofos/Evaluating-the-Implicit-Midpoint-Integrator}.

\subsection{Summary of Integrators}

We consider two variants of the generalized leapfrog method and two variants of the implicit midpoint integrator, which we summarily describe as follows.
\begin{description}
\itemsep0em
\item[G.L.F.(a)] An implementation of the generalized leapfrog integrator as presented in \cref{alg:generalized-leapfrog}.
\item[G.L.F.(b)] An implementation of the generalized leapfrog integrator that caches repeated calculations and which is specific to Hamiltonians in the form of \cref{eq:hamiltonian-energy}. See \cref{alg:smart-generalized-leapfrog} in \cref{app:implementation-of-integrators}. G.L.F.(b) is mathematically identical to G.L.F.(a), but this implementation avoids some redundant computation. Differences between the outputs of G.L.F.(a) and G.L.F.(b) are due to random seeds and machine error in computation.
\item[I.M.(a)] An implementation of the implicit midpoint integrator as presented in \cref{alg:implicit-midpoint}.
\item[I.M.(b)] An implementation of the implicit midpoint integrator that implicitly computes the midpoint followed by an explicit Euler step, as advocated by \cite{leimkuhler_reich_2005}. See \cref{alg:smart-implicit-midpoint} in \cref{app:implementation-of-integrators}.
\end{description}
In all of our implementations, we use fixed point iterations in order to find solutions to implicitly-defined relations. This is the approach advocated by \citet{Hairer:1250576}. 
Additional details are presented in \cref{app:implementation-of-integrators}.

\subsection{Banana-Shaped Distribution}\label{subsec:experiment-banana-shaped}

\begin{table*}[t!]
    \centering
    \scriptsize
    \begin{tabular}{lll|rrrrrr}
\toprule
     &      &                            &  Acc. Prob. & Time (Sec.) &   Mean ESS &    Min. ESS &  Mean ESS / Sec. &  Min. ESS / Sec. \\
Step Size & Num. Steps & Method &          &            &            &               &              \\
\midrule
0.1 & 5 & G.L.F.(a) & $0.62 \pm 0.01$ & $400.33 \pm 6.83$ & $486.32 \pm 17.89$ & $286.28 \pm 16.09$ & $1.21 \pm 0.04$ & $0.71 \pm 0.04$ \\
& & G.L.F.(b) & $0.61 \pm 0.01$ & $145.99 \pm 2.20$ & $491.07 \pm 22.03$ & $301.90 \pm 16.21$ & $3.36 \pm 0.13$ & $2.07 \pm 0.10$ \\
& & I.M.(a) & $0.98 \pm 0.00$ & $102.37 \pm 1.13$ & $884.09 \pm 27.39$ & $620.26 \pm 30.08$ & $8.65 \pm 0.31$ & $6.07 \pm 0.32$ \\
& & I.M.(b) & $0.98 \pm 0.00$ & $95.28 \pm 2.28$ & $857.61 \pm 24.99$ & $619.91 \pm 27.09$ & $9.03 \pm 0.28$ & $6.53 \pm 0.31$ \\
& 10 & G.L.F.(a) & $0.50 \pm 0.01$ & $615.14 \pm 6.51$ & $1038.47 \pm 25.68$ & $778.80 \pm 30.92$ & $1.69 \pm 0.05$ & $1.27 \pm 0.05$ \\
& & G.L.F.(b) & $0.49 \pm 0.01$ & $231.55 \pm 2.48$ & $1027.78 \pm 25.19$ & $782.85 \pm 14.87$ & $4.44 \pm 0.12$ & $3.38 \pm 0.07$ \\
& & I.M.(a) & $0.98 \pm 0.00$ & $194.82 \pm 2.39$ & $3018.50 \pm 70.23$ & $2518.65 \pm 83.19$ & $15.51 \pm 0.39$ & $12.94 \pm 0.43$ \\
& & I.M.(b) & $0.98 \pm 0.00$ & $172.12 \pm 1.45$ & $3025.14 \pm 55.19$ & $2540.88 \pm 99.95$ & $17.58 \pm 0.34$ & $14.76 \pm 0.56$ \\
& 50 & G.L.F.(a) & $0.13 \pm 0.00$ & $2133.63 \pm 64.84$ & $192.82 \pm 24.27$ & $79.12 \pm 14.55$ & $0.09 \pm 0.01$ & $0.04 \pm 0.01$ \\
& & G.L.F.(b) & $0.14 \pm 0.00$ & $786.40 \pm 18.50$ & $247.58 \pm 28.90$ & $119.23 \pm 18.68$ & $0.31 \pm 0.03$ & $0.15 \pm 0.02$ \\
& & I.M.(a) & $0.95 \pm 0.00$ & $938.79 \pm 15.17$ & $4173.70 \pm 199.89$ & $3207.59 \pm 113.94$ & $4.47 \pm 0.25$ & $3.43 \pm 0.14$ \\
& & I.M.(b) & $0.95 \pm 0.00$ & $834.63 \pm 13.42$ & $3928.27 \pm 159.13$ & $3158.40 \pm 93.76$ & $4.73 \pm 0.24$ & $3.80 \pm 0.15$ \\
\bottomrule
\end{tabular}
    \caption{Comparison of the implicit midpoint and generalized leapfrog integrators on sampling from the banana-shaped distribution. To assess performance of the sampler, we measure the effective sample size (ESS) and present per-second timing comparisons for the mean and minimum ESS. The hypothesis that the implicit midpoint integrator should exhibit better energy conservation is captured in the acceptance probability of the Markov chain. Results are averaged over ten trials.}
    \label{tab:banana-ess}
\end{table*}
\begin{figure*}[t!]
    \centering
    \begin{subfigure}[b]{0.3\textwidth}
        \centering
        \includegraphics[width=\textwidth]{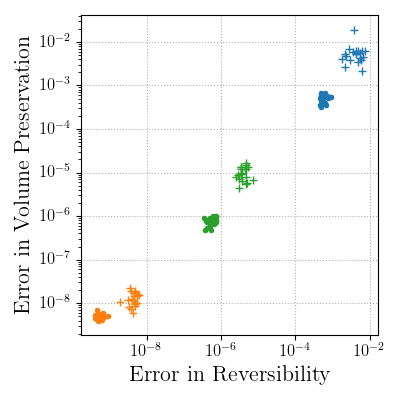}
        \caption{$\epsilon=1/10$ and $5$ steps}
    \end{subfigure}
    ~
    \begin{subfigure}[b]{0.3\textwidth}
        \centering
        \includegraphics[width=\textwidth]{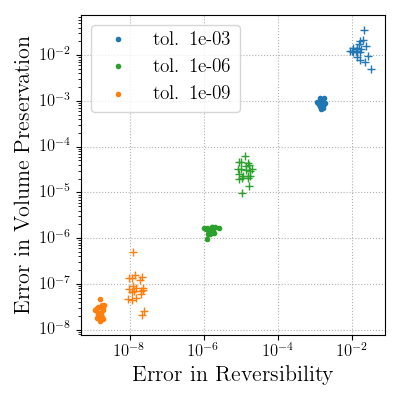}
        \caption{$\epsilon=1/10$ and $10$ steps}
    \end{subfigure}
    ~
    \begin{subfigure}[b]{0.3\textwidth}
        \centering
        \includegraphics[width=\textwidth]{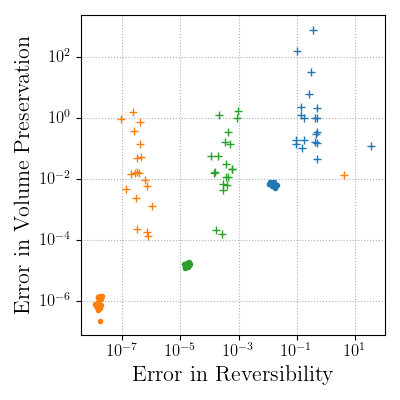}
        \caption{$\epsilon=1/10$ and $50$ steps}
    \end{subfigure}
    \caption{Comparison between the error in symmetry and error in volume preservation properties of the implicit midpoint integrator and the generalized leapfrog integrator on the banana-shaped distribution. We observe that the implicit midpoint integrator tends to produce transitions whose median reversibility and volume preservation can be an order of magnitude, or more, better than the generalized leapfrog integrator. Each point is the median of one-hundred measurements of symmetry and reversibility shown for each of the ten trials. For $\epsilon=1/10$ and fifty integration steps, the generalized leapfrog integrator exhibits severely divergent behavior. {\bf The implicit midpoint is represented by the symbol $(\boldsymbol{\cdot})$ and the generalized leapfrog by the symbol $(+)$.}}
    \label{fig:banana-symmetry-volume-preservation}
\end{figure*}

The banana-shaped distribution was proposed in a discussion to \cite{doi:10.1111/j.1467-9868.2010.00765.x} as a representative example of the ridge-like posterior structure that can manifest in non-identifiable models. The banana-shaped distribution is defined by the following generative model.
\begin{align}
    y_i \vert\theta_1,\theta_2 &\sim \mathrm{Normal}(\theta_1 + \theta_2^2, \sigma_y^2) ~~~\mathrm{for}~i=1,\ldots, n \\
    \theta_i &\sim \mathrm{Normal}(0, \sigma_\theta^2) ~~~~~~~~~~~~~~\mathrm{for}~i\in\set{1,2}.
\end{align}
For the banana-shaped distribution, the Riemannian metric is
\begin{align}
  \mathbb{G}(\theta_1,\theta_2) = \begin{pmatrix} \frac{n}{\sigma^2_y} + \frac{1}{\sigma^2_\theta} & \frac{2n\theta_2}{\sigma_y^2} \\ \frac{2n\theta_2}{\sigma_y^2} & \frac{4n\theta_2^2}{\sigma_y^2} + \frac{1}{\sigma_\theta^2} \end{pmatrix}.
\end{align}
In our experiments, we take $n=100$. We generate observations $\set{y_1,\ldots,y_{100}}$ from the banana-shaped distribution by setting $\theta_1 = 1/2$, $\theta_2=1/\sqrt{2}$, and $\sigma_y = \sigma_\theta = 2$. We then attempt to sample the posterior distribution of $(\theta_1,\theta_2)$ using RMHMC when integration is performed using the implicit midpoint algorithm or the generalized leapfrog method. We consider two step-sizes $\set{0.01, 0.1}$ and a number of integration steps in $\set{5, 10, 50}$. We attempt to draw 10,000 samples from the posterior. Each of these configurations is replicated ten times.

Results are shown in \cref{tab:banana-ess}, demonstrating that the I.M.(a) and (b) integrators are able to maintain high energy conservation at step-sizes for which the G.L.F.(a) and (b) variants cannot. As a consequence, Markov chains using I.M.(a) or I.M.(b) are able to achieve very high effective sample sizes (ESS); moreover, because the cost of evaluating the gradients of the banana-shaped posterior is not too large, these Marko chains also exhibits superior performance {\it on the timing comparisons}. We find that I.M.(a) and I.M.(b) perform similarly.
In addition to energy conservation, an essential component of HMC are volume preservation and reversibility (recall properties (i) and (ii) from \cref{sec:introduction}). Using the samples drawn by the Markov chains with either integrator, we may compute numerical estimates of the degree to which these properties are satisfied. 
We give a detailed description of the volume preservation and reversibility metrics in \cref{app:volume-preservation-and-symmetric-metrics}.
We use one-hundred randomly selected samples generated from the Markov chains in order to compute these statistics. Results showing the {\it median} reversibility versus the {\it median} difference from unit Jacobian are shown in \cref{fig:banana-symmetry-volume-preservation}. These results show that the median symmetry and volume preservation of the implicit midpoint integrator is approximately an order of magnitude more faithfully preserved than is the case for the generalized leapfrog method.

\subsection{Hierarchical Neal's Funnel Distribution}

\begin{table*}[t!]
    \centering
    \scriptsize
    \begin{tabular}{lll|rrrrrr}
\toprule
     &      &                            &  Acc. Prob. & Time (Sec.) &   Mean ESS &    Min. ESS &  Mean ESS / Sec. &  Min. ESS / Sec. \\
Num. Steps & Step Size & Method &          &            &            &               &              \\
\midrule
20 & 0.1 & G.L.F.(a) & $0.99 \pm 0.00$ & $1317.57 \pm 16.89$ & $15614.52 \pm 177.89$ & $340.56 \pm 17.41$ & $11.86 \pm 0.16$ & $0.26 \pm 0.01$ \\
& & I.M.(a) & $1.00 \pm 0.00$ & $1315.24 \pm 57.79$ & $16199.00 \pm 157.60$ & $383.48 \pm 22.16$ & $12.47 \pm 0.44$ & $0.30 \pm 0.02$ \\
& & I.M.(b) & $1.00 \pm 0.00$ & $1147.45 \pm 64.92$ & $15697.78 \pm 148.81$ & $392.22 \pm 23.11$ & $14.02 \pm 0.74$ & $0.35 \pm 0.03$ \\
& 0.2 & G.L.F.(a) & $0.96 \pm 0.00$ & $1933.64 \pm 97.74$ & $29457.72 \pm 353.68$ & $1511.75 \pm 28.61$ & $15.55 \pm 0.79$ & $0.79 \pm 0.03$ \\
& & I.M.(a) & $0.99 \pm 0.00$ & $1669.69 \pm 78.93$ & $32055.29 \pm 257.52$ & $1588.73 \pm 54.01$ & $19.52 \pm 0.86$ & $0.96 \pm 0.05$ \\
& & I.M.(b) & $0.99 \pm 0.00$ & $1441.81 \pm 61.06$ & $31585.28 \pm 197.07$ & $1558.54 \pm 36.83$ & $22.17 \pm 0.79$ & $1.10 \pm 0.05$ \\
& 0.5 & G.L.F.(a) & $0.36 \pm 0.00$ & $2389.27 \pm 49.42$ & $2144.93 \pm 103.34$ & $1771.92 \pm 140.71$ & $0.90 \pm 0.05$ & $0.74 \pm 0.06$ \\
& & I.M.(a) & $0.85 \pm 0.00$ & $2981.80 \pm 71.09$ & $10984.29 \pm 123.63$ & $10147.33 \pm 121.32$ & $3.70 \pm 0.09$ & $3.42 \pm 0.07$ \\
& & I.M.(b) & $0.85 \pm 0.00$ & $2728.61 \pm 82.50$ & $10563.94 \pm 210.16$ & $9711.29 \pm 264.30$ & $3.90 \pm 0.13$ & $3.58 \pm 0.13$ \\
\bottomrule
\end{tabular}
    \caption{Comparison of the implicit midpoint and the naive generalized leapfrog integrators on sampling from Neal's funnel distribution. We see that the implicit midpoint integrator is able to take large steps and produce an effective sample size that outperforms the generalized leapfrog integrator even in the time-normalized performance.}
    \label{tab:neal-funnel-ess}
\end{table*}
\begin{figure*}[t!]
    \centering
    \begin{subfigure}[b]{0.3\textwidth}
        \centering
        \includegraphics[width=\textwidth]{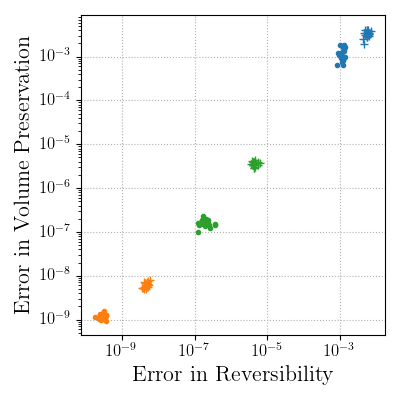}
        \caption{$\epsilon=1/10$}
    \end{subfigure}
    ~
    \begin{subfigure}[b]{0.3\textwidth}
        \centering
        \includegraphics[width=\textwidth]{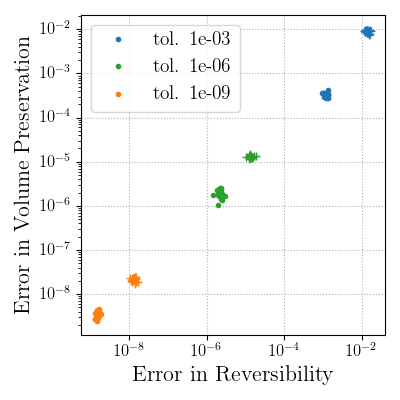}
        \caption{$\epsilon=1/5$}
    \end{subfigure}
    ~
    \begin{subfigure}[b]{0.3\textwidth}
        \centering
        \includegraphics[width=\textwidth]{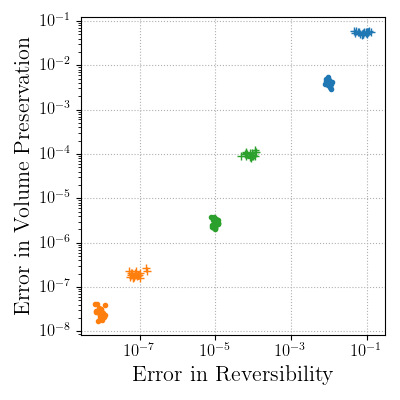}
        \caption{$\epsilon=1/2$}
    \end{subfigure}
    \caption{Comparison of the degree to which the implicit midpoint and generalized leapfrog integrators violate reversibility and volume preservation on Neal's funnel distribution. We observe that the implicit midpoint integrator exhibits better symmetry and volume preservation. {\bf The implicit midpoint is represented by the symbol $(\boldsymbol{\cdot})$ and the generalized leapfrog by the symbol $(+)$.}}
    \label{fig:neal-funnel-symmetry-volume}
\end{figure*}

As an example of a hierarchical Bayesian posterior, we consider Neal's funnel distribution defined by,
\begin{align}
    x_i &\sim \mathrm{Normal}(0, \exp(-v))~~\mathrm{for}~i=1,\ldots,10 \\
    v &\sim \mathrm{Normal}(0, 9).
\end{align}
Due to the hierarchical structure of the distribution, the Hessian of the distribution is not convex and therefore cannot be used to construct a Riemannian metric on its own. Instead, we follow the approach proposed in \citet{softabs} and adopt the SoftAbs transformation of the Hessian in order to construct a positive definite Riemannian metric. This allows us to sample all variables of the hierarchical distribution jointly. For RMHMC, we consider an integration step-size in $\set{0.1, 0.2, 0.5}$ and we attempt to draw 10,000 samples of $(x_1,\ldots,x_{10}, v)$ from Neal's funnel distribution.

Results are presented in \cref{tab:neal-funnel-ess}. For the largest step-size, the I.M.(a) and (b) integrators are able to maintain high acceptance probabilities. Markov chains using the I.M.(a) or (b) method achieve the best {\it minimum} ESS per-second. This example presents a circumstance wherein the time-normalized best-case performance of mean and minimum effective sample sizes did not co-occur in the same parameter configuration. Nevertheless, when optimizing for the highest {\it mean} ESS per-second, the I.M.(a) and (b) methods also outperform the G.L.F.(a) integrator. We visualize the symmetry and volume preservation in \cref{fig:neal-funnel-symmetry-volume}; the implicit midpoint integrator exhibits better symmetry and volume preservation for the same convergence criterion.

\subsection{Stochastic Volatility Model}

\begin{table*}[t!]
    \centering
    \tiny
    \begin{tabular}{l|rrrrrr|rr|rr}
\toprule
       &&&&&&& \multicolumn{2}{c}{Volume Preservation} & \multicolumn{2}{c}{Symmetry} \\
    Method   & Acc. Prob. & Time (Sec.) &   Mean ESS &    Min. ESS &  \begin{tabular}[c]{@{}l@{}}Mean ESS \\ (Sec.)\end{tabular} &  \begin{tabular}[c]{@{}l@{}}Min. ESS \\ (Sec.)\end{tabular} & Median &    $90^\mathrm{th}$-Per. &  Median &    $90^\mathrm{th}$-Per. \\
\midrule
G.L.F.(a) & $0.78 \pm 0.0$ & $2616.70 \pm 16.5$ & $294.51 \pm 10.7$ & $121.79 \pm 7.7$ & $0.11 \pm 0.0$ & $0.05 \pm 0.0$ & $7.2\mathrm{e-}07$ & $3.1\mathrm{e-}06$ & $4.9\mathrm{e-}06$ & $2.5\mathrm{e-}05$ \\
G.L.F.(b) & $0.78 \pm 0.0$ & $2339.78 \pm 9.5$ & $297.82 \pm 9.2$ & $132.79 \pm 6.6$ & $0.13 \pm 0.0$ & $0.06 \pm 0.0$ & $6.7\mathrm{e-}07$ & $2.8\mathrm{e-}06$ & $5.0\mathrm{e-}06$ & $2.4\mathrm{e-}05$ \\
I.M.(a) & $0.80 \pm 0.0$ & $2761.61 \pm 8.3$ & $309.10 \pm 10.5$ & $133.32 \pm 7.8$ & $0.11 \pm 0.0$ & $0.05 \pm 0.0$ & $8.7\mathrm{e-}08$ & $2.2\mathrm{e-}07$ & $1.3\mathrm{e-}06$ & $2.4\mathrm{e-}06$ \\
I.M.(b) & $0.80 \pm 0.0$ & $2710.47 \pm 8.4$ & $296.72 \pm 9.2$ & $127.35 \pm 7.3$ & $0.11 \pm 0.0$ & $0.05 \pm 0.0$ & $1.7\mathrm{e-}07$ & $4.5\mathrm{e-}07$ & $2.8\mathrm{e-}06$ & $4.7\mathrm{e-}06$ \\
\bottomrule
\end{tabular}
    \caption{Comparison of the implicit midpoint generalized leapfrog integrators on the stochastic volatility model. We see that the implicit midpoint integrator is competitive on the effective sample size and timing comparisons. We also evaluate the median reversibility and volume preservation of the implicit midpoint and the generalized leapfrog integrators on the stochastic volatility model with $\delta = 1\times10^{-6}$. Here we see that the implicit midpoint integrator enjoys better conservation of volume and reversibility.}
    \label{tab:stochastic-volatility-symmetry-volume-preservation}
\end{table*}

While Neal's funnel is a hierarchical distribution, it is not sampled in a hierarchical manner, instead sampling all variables jointly using the SoftAbs Riemannian metric \cite{softabs}. Here, we consider a stochastic volatility model whose posterior includes the stochastic volatilities as well as latent hyperparameters of the model; we will sample these variables using an alternating Gibbs procedure. Following \citet{doi:10.1111/j.1467-9868.2010.00765.x}, the stochastic volatility model is defined, for $t=1, \ldots,T$, by,
\begin{align}
    y_t \vert\beta, x_t &\sim \mathrm{Normal}(0, \beta^2e^{x_t}) \\
    x_t\vert x_{t-1}, \phi, \sigma^2 &\sim \mathrm{Normal}(\phi x_{t-1}, \sigma^2)
\end{align}
where $x_1 \sim\mathrm{Normal}(0, \sigma^2 / (1-\phi^2))$, $\pi(\beta) \propto 1/\beta$, $\sigma^2 \sim\mathrm{Inv-}\chi^2(10, 0.05)$, and $(\phi + 1)/ 2\sim\mathrm{Beta}(20, 1.5)$. The sampler proceeds by alternating between sampling the conditional posteriors of $(x_1,\ldots, x_T)\vert (y_1,\ldots,y_T), \phi, \beta, \sigma^2$ and $\phi, \beta,\sigma^2\vert (x_1,\ldots, x_T), (y_1,\ldots,y_T)$. The Riemannian metric of this first posterior is constant with respect to $(x_1,\ldots, x_T)$; therefore, sampling is carried out using the standard leapfrog integrator. The second distribution has a position-dependent Riemannian metric, necessitating the use of implicitly-defined integrators; here, we compare the implicit midpoint and generalized leapfrog integrators. For details of the Riemannian structures of the conditional posteriors, see \citet{doi:10.1111/j.1467-9868.2010.00765.x}.

In our experiments, we set $T=1,000$ and use fifty integration steps with a step-size of $0.1$ to sample $(x_1,\ldots, x_T)\vert (y_1,\ldots,y_T), \phi, \beta, \sigma^2$ and six integration steps with a step-size of $0.5$ to sample $\phi, \beta,\sigma^2\vert (x_1,\ldots, x_T), (y_1,\ldots,y_T)$. We seek to sample $20,000$ times from the posterior and use a burn-in period of $10,000$ iterations. We repeat this experiment one-hundred times for each integrator. Effective sample size metrics and measures of the volume preservation and symmetry are are presented in \cref{tab:stochastic-volatility-symmetry-volume-preservation} {\it for the parameters $\phi$, $\beta$, and $\sigma^2$}. We find that the I.M.(a) and (b) integrators are comparable to the G.L.F.(a) and (b) methods in terms of their time-normalized performance. However, volume preservation and symmetry are better for the I.M.(a) and (b) integrators.

\subsection{Fitzhugh-Nagumo ODE Model}

\begin{figure*}[t!]
    \centering
    \begin{subfigure}[b]{0.3\textwidth}
        \centering
        \includegraphics[width=\textwidth]{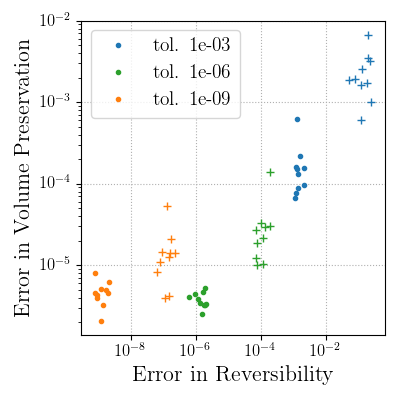}
        \caption{One integration step}
    \end{subfigure}
    ~
    \begin{subfigure}[b]{0.3\textwidth}
        \centering
        \includegraphics[width=\textwidth]{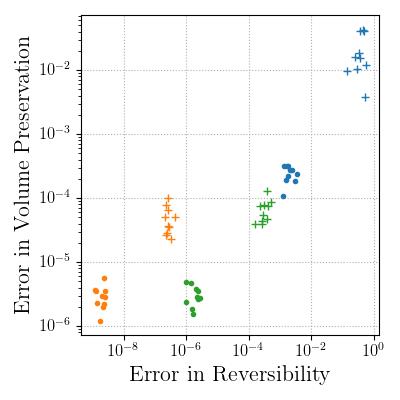}
        \caption{Two integration steps}
    \end{subfigure}
    ~
    \begin{subfigure}[b]{0.3\textwidth}
        \centering
        \includegraphics[width=\textwidth]{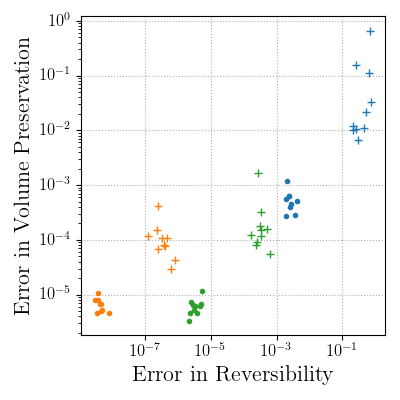}
        \caption{Five integration steps}
    \end{subfigure}
    \caption{Comparison of the degree to which the implicit midpoint and generalized leapfrog integrators violate reversibility and volume preservation on the Fitzhugh-Nagumo posterior. We observe that for the chosen convergence tolerances, the implicit midpoint integrator tends to exhibit better reversibility and volume preservation except for the smallest tolerance where the generalized leapfrog has slightly better volume preservation but worse reversibility. {\bf The implicit midpoint is represented by the symbol $(\boldsymbol{\cdot})$ and the generalized leapfrog by the symbol $(+)$.}}
    \label{fig:fitzhugh-nagumo-symmetry-volume}
\end{figure*}
\begin{table*}[t!]
    \centering
    \scriptsize
    \begin{tabular}{lll|rrrrrr}
\toprule
     &      &                            &  Acc. Prob. & Time (Sec.) &   Mean ESS &    Min. ESS &  Mean ESS / Sec. &  Min. ESS / Sec. \\
Step Size & Num. Steps & Method &          &            &            &               &              \\
\midrule
1.0 & 1 & G.L.F.(a) & $0.74 \pm 0.01$ & $4441.16 \pm 117.73$ & $276.39 \pm 13.48$ & $235.39 \pm 10.55$ & $0.06 \pm 0.00$ & $0.05 \pm 0.00$ \\
& & G.L.F.(b) & $0.73 \pm 0.01$ & $1343.63 \pm 76.12$ & $272.72 \pm 7.96$ & $235.52 \pm 8.99$ & $0.21 \pm 0.01$ & $0.18 \pm 0.01$ \\
& & I.M.(a) & $0.95 \pm 0.00$ & $4523.49 \pm 160.00$ & $227.01 \pm 13.47$ & $199.10 \pm 13.70$ & $0.05 \pm 0.00$ & $0.04 \pm 0.00$ \\
& & I.M.(b) & $0.95 \pm 0.00$ & $4486.29 \pm 147.33$ & $248.11 \pm 8.74$ & $225.67 \pm 9.11$ & $0.06 \pm 0.00$ & $0.05 \pm 0.00$ \\
& 2 & G.L.F.(a) & $0.74 \pm 0.01$ & $8100.90 \pm 423.17$ & $1089.36 \pm 55.17$ & $907.30 \pm 75.02$ & $0.14 \pm 0.01$ & $0.11 \pm 0.01$ \\
& & G.L.F.(b) & $0.75 \pm 0.01$ & $1912.73 \pm 48.99$ & $1109.78 \pm 71.50$ & $917.52 \pm 98.66$ & $0.58 \pm 0.04$ & $0.48 \pm 0.05$ \\
& & I.M.(a) & $0.94 \pm 0.00$ & $8704.37 \pm 250.22$ & $1387.96 \pm 40.53$ & $1212.91 \pm 42.68$ & $0.16 \pm 0.01$ & $0.14 \pm 0.01$ \\
& & I.M.(b) & $0.94 \pm 0.00$ & $8423.40 \pm 139.60$ & $1366.07 \pm 56.59$ & $1213.01 \pm 55.18$ & $0.16 \pm 0.00$ & $0.14 \pm 0.01$ \\
& 5 & G.L.F.(a) & $0.70 \pm 0.01$ & $20096.64 \pm 384.16$ & $188.59 \pm 27.78$ & $90.00 \pm 19.15$ & $0.01 \pm 0.00$ & $0.00 \pm 0.00$ \\
& & G.L.F.(b) & $0.71 \pm 0.01$ & $5138.31 \pm 300.83$ & $150.03 \pm 23.06$ & $102.32 \pm 22.10$ & $0.03 \pm 0.00$ & $0.02 \pm 0.00$ \\
& & I.M.(a) & $0.94 \pm 0.00$ & $22069.48 \pm 650.03$ & $954.94 \pm 93.17$ & $818.16 \pm 91.14$ & $0.04 \pm 0.01$ & $0.04 \pm 0.01$ \\
& & I.M.(b) & $0.94 \pm 0.00$ & $20575.78 \pm 506.21$ & $1095.88 \pm 130.14$ & $821.77 \pm 80.15$ & $0.05 \pm 0.01$ & $0.04 \pm 0.00$ \\
\bottomrule
\end{tabular}
    \caption{Comparison of the implicit midpoint and generalized leapfrog integrators on sampling from the posterior of the Fitzhugh-Nagumo model. In this example, the higher acceptance probability of the implicit midpoint integrator did not produce a performance increase for a single integration step. The implicit midpoint integrator becomes super-efficient in the two-step regime, but cannot compete with the generalized leapfrog integrator's computational advantages.}
    \label{tab:fitzhugh-nagumo-ess}
\end{table*}

The Fitzhugh-Nagumo ordinary differential equation is a model of neural spiking activity. It is described by two time-varying measurements whose dynamics obey,
\begin{align}
    \label{eq:fitzhugh-nagumo-v} \dot{v} &= \theta_3\paren{v - \frac{v^3}{3} + r} \\
    \label{eq:fitzhugh-nagumo-r} \dot{r} &= -\paren{\frac{v-\theta_1+\theta_2r}{\theta_3}}.
\end{align}
Consider the setting wherein one has 200 observations of the Fitzhugh-Nagumo dynamics at equally-spaced times between zero and ten. Assume moreover that these observations have been corrupted by i.i.d. Gaussian noise with a known standard deviation of $\sigma = 1/2$. If we equip the parameters $\theta_1$, $\theta_2$, and $\theta_3$ with standard normal priors, we may use the dynamics of \cref{eq:fitzhugh-nagumo-v,eq:fitzhugh-nagumo-r} and the assumed noise distribution in order to sample the posterior of $(\theta_1, \theta_2, \theta_3)$. Let $(v_n, r_n)$ be the solution of the Fitzhugh-Nagumo ODE at the $n^\text{th}$ time period. For the Fitzhugh-Nagumo differential equation model, the $(i,j)$-entry of the metric is,
\begin{align}
  \mathbb{G}_{ij}(\theta_1,\theta_2,\theta_3) = \frac{1}{\sigma^2} \paren{\sum_{n=1}^{200} \frac{\partial v_n}{\partial \theta_i} \frac{\partial v_n}{\partial \theta_j} + \frac{\partial r_n}{\partial \theta_i} \frac{\partial r_n}{\partial \theta_j}} + \delta_{ij}.
\end{align}
In generating data from the Fitzhugh-Nagumo model, we set $\theta_1=\theta_2=0.2$ and $\theta_3=3$. The dynamics are integrated using SciPy's \texttt{odeint} function and gradients are {\it approximated} by forward sensitivity analysis as in \citet{doi:10.1111/j.1467-9868.2010.00765.x}. We consider an integration step-size of $\epsilon=1$ and a number of integration steps in $\set{1, 2, 5}$; each configuration is replicated ten times. We sample 1,000 times from the posterior.

We expect the Fitzhugh-Nagumo ODE model to favor the generalized leapfrog integrator because of the complexity of evaluating the log-posterior, the gradient of the log-posterior, the Riemannian metric, and the gradient of the Riemannian metric, each of which involves solving a system of differential equations. Therefore, the caching behavior associated to the G.L.F.(b) integrator gives it an advantage here. \Cref{tab:fitzhugh-nagumo-ess} shows the results of inferences in the Fitzhugh-Nagumo posterior. We observe that for a single-step, the I.M.(a) and (b) integrators appears to perform somewhat worse than the G.L.F.(a) and (b) variants, even on the measures of ESS that ignore timing; this occurs despite the larger acceptance rate enjoyed by the implicit midpoint integrator. For two integration steps, the inferences produced by I.M.(a) and (b) become super-efficient; however, G.L.F.(a) and (b) are also efficient and the computational advantage of the (b) variant cause it to have superior performance in the timing metrics. For the largest number of steps, the performance of G.L.F.(a) and (b) deteriorates so that the I.M.(a) and (b) integrators outperform them even on the timing comparison.

We also evaluate the degree to which the numerical integrator possesses the properties of symmetry and volume preservation. The results are shown in \cref{fig:fitzhugh-nagumo-symmetry-volume}. We see that the implicit midpoint integrator offers a clear advantage in numerical symmetry, and performs better on volume preservation as well. 

\section{Conclusion}

This work has considered the implicit midpoint integrator as a substitute for the generalized leapfrog integrator for use in RMHMC. Inspired by the theory of the conserved quantities of numerical integrators, we hypothesized that the implicit midpoint integrator would have better energy conservation in posterior distributions that are approximately Gaussian. Hamiltonian Monte Carlo requires that its integrators are volume preserving and reversible; we give numerical assessments of the extent to which these properties are present in implementations of these integrators, which differ from their theoretical representation when a convergence tolerance is used to halt a fixed point iteration. We find that the implicit midpoint integrator has superior energy conservation, conservation of volume, and reversibility across several Bayesian inference tasks. In three of the four example applications, the implicit midpoint integrator met or exceeded the {\it time-normalized} performance of the generalized leapfrog integrator. This, combined with its better volume preservation and reversibility, leads us to conclude that it is a method worth consideration when implementing RMHMC. 


\subsection*{Acknowledgments}

The authors would like to thank Marcus A. Brubaker for helpful discussions.

This material is based upon work supported by the National Science Foundation Graduate Research Fellowship under Grant No. 1752134. Any opinion, findings, and conclusions or recommendations expressed in this material are those of the authors(s) and do not necessarily reflect the views of the National Science Foundation. RRL was supported in part by NIH/NIGMS 1R01GM136780-01.

\bibliography{thebib}
\bibliographystyle{icml2021}

\newpage
\onecolumn
\appendix

\newpage
\section{Momentum Negation Symmetry of Implicit Midpoint}\label{app:momentum-negation-implicit-midpoint}

\begin{lemma}\label{lem:implicit-midpoint-momentum-flip}
Given a Hamiltonian $H$ in the form of \cref{eq:hamiltonian-energy}, step-size $\epsilon$, and initial position $(q, p)$, compute $(q', p')$ according to \cref{alg:implicit-midpoint} with $\delta=0$. If one then computes $(q'', p'')$ from initial position $(q', -p')$ using \cref{alg:implicit-midpoint} a second time (with the same Hamiltonian, step-size, and $\delta=0$), then $q''=q$ and $-p'' = p$.
\end{lemma}
\Cref{lem:implicit-midpoint-momentum-flip} establishes that the implicit midpoint integrator is suitable for HMC in that it satisfies properties (i) and (ii).

\begin{proof}
Consider the initial condition $(q, p)$ and a fixed step-size of $\epsilon$. For the Riemannian manifold Hamiltonian Monte Carlo, the implicit midpoint integrator computes the following updates:
\begin{align}
    \label{eq:implicit-reverse-position} q'_i &= q_i + \epsilon\paren{\sum_{j=1}^m \mathbb{G}^{-1}_{ij}\paren{\frac{q' + q}{2}}\paren{\frac{p'_j + p_j}{2}}} \\
    \label{eq:implicit-reverse-momentum} \begin{split}
    p'_i &= p_i + \epsilon\left(-\frac{\partial}{\partial q_i}\mathcal{L}\paren{\frac{q' + q}{2}} - \frac{1}{2} \mathrm{trace}\paren{\mathbb{G}^{-1}\paren{\frac{q' + q}{2}} \frac{\partial}{\partial q_i} \mathbb{G}\paren{\frac{q' + q}{2}}} + \right.\\
    &\qquad \left.\frac{1}{2} \paren{\frac{p' + p}{2}}^\top\mathbb{G}^{-1}\paren{\frac{q' + q}{2}} \frac{\partial}{\partial q_i}\mathbb{G}\paren{\frac{q' + q}{2}} \mathbb{G}^{-1}\paren{\frac{q' + q}{2}} \paren{\frac{p' + p}{2}}\right)
    \end{split}
\end{align}
What we want to show is that if we compute $(q', p')$, negate the momentum $(q', p')\mapsto (q', -p')$, and apply the implicit midpoint integrator a second time, then we arrive at $(q, -p)$. Thus, we need to establish that $(q, -p)$ is a fixed point of the relations,
\begin{align}
    q''_i &= q'_i + \epsilon\paren{\sum_{j=1}^m \mathbb{G}^{-1}_{ij}\paren{\frac{q'' + q'}{2}}\paren{\frac{p''_j + (-p_j')}{2}}} \\
    \begin{split}
    p''_i &= -p'_i + \epsilon\left(-\frac{\partial}{\partial q_i}\mathcal{L}\paren{\frac{q'' + q'}{2}} - \frac{1}{2} \mathrm{trace}\paren{\mathbb{G}^{-1}\paren{\frac{q'' + q'}{2}} \frac{\partial}{\partial q_i} \mathbb{G}\paren{\frac{q'' + q'}{2}}} + \right.\\
    &\qquad \left.\frac{1}{2} \paren{\frac{p'' + (-p')}{2}}^\top\mathbb{G}^{-1}\paren{\frac{q'' + q'}{2}} \frac{\partial}{\partial q_i}\mathbb{G}\paren{\frac{q'' + q'}{2}} \mathbb{G}^{-1}\paren{\frac{q'' + q'}{2}} \paren{\frac{p'' + (-p')}{2}}\right).
    \end{split}
\end{align}
Plugging in we obtain,
\begin{align}
    q'_i + \epsilon\paren{\sum_{j=1}^m \mathbb{G}^{-1}_{ij}\paren{\frac{q + q'}{2}}\paren{\frac{(-p_j) + (-p_j')}{2}}} &= q'_i - \epsilon\paren{\sum_{j=1}^m \mathbb{G}^{-1}_{ij}\paren{\frac{q + q'}{2}}\paren{\frac{p_j + p_j'}{2}}} \\
    &= q_i
\end{align}
by rearranging \cref{eq:implicit-reverse-position}. For notational simplicity let us define
\begin{align}
    U\paren{q} &\defeq -\frac{\partial}{\partial q_i}\mathcal{L}\paren{q} - \frac{1}{2} \mathrm{trace}\paren{\mathbb{G}^{-1}\paren{q} \frac{\partial}{\partial q_i} \mathbb{G}\paren{q}} \\
    R\paren{q} &\defeq \frac{1}{2}\mathbb{G}^{-1}\paren{q} \frac{\partial}{\partial q_i}\mathbb{G}\paren{q} \mathbb{G}^{-1}\paren{q}
\end{align}
so that
\begin{align}
    p''_i = -p'_i + \epsilon \paren{U\paren{\frac{q'' + q}{2}} + \paren{\frac{p'' + (-p')}{2}}^\top R\paren{\frac{q'' + q}{2}}\paren{\frac{p'' + (-p')}{2}}}.
\end{align}
Plugging in, we obtain,
\begin{align}
    & -p'_i + \epsilon \paren{U\paren{\frac{q + q}{2}} + \paren{\frac{(-p) + (-p')}{2}}^\top R\paren{\frac{q + q}{2}}\paren{\frac{(-p) + (-p')}{2}}} \\
    =& -p'_i + \epsilon \paren{U\paren{\frac{q + q}{2}} + \paren{\frac{p + p'}{2}}^\top R\paren{\frac{q + q}{2}}\paren{\frac{p + p'}{2}}} \\
    =& -p_i
\end{align}
which follows from negating \cref{eq:implicit-reverse-momentum} and rearranging.
\end{proof}
\newpage
\section{Implicit Midpoint Eigenvalues}\label{app:implicit-midpoint-eigenvalues}

Let $z=(q,p)$ and consider a quadratic Hamiltonian of the form,
\begin{align}
    H(z) &= z^\top \mathbf{A}z \\
    &= \begin{pmatrix} q^\top & p^\top \end{pmatrix} \begin{pmatrix} \Sigma & 0 \\ 0 & \Sigma \end{pmatrix} \begin{pmatrix} q \\ p \end{pmatrix}.
\end{align}
The associated Hamiltonian vector field is,
\begin{align}
    \dot{z} &= \begin{pmatrix} \dot{q} \\ \dot{p} \end{pmatrix} \\
    &= \begin{pmatrix} 0 & \mathrm{Id} \\ -\mathrm{Id} & 0 \end{pmatrix} \begin{pmatrix} \Sigma & 0 \\ 0 & \Sigma \end{pmatrix} \begin{pmatrix} q \\ p \end{pmatrix} \\
    &= \underbrace{\begin{pmatrix} 0 & \Sigma \\ -\Sigma & 0 \end{pmatrix}}_{\mathbf{J}} \begin{pmatrix} q \\ p \end{pmatrix} \\
    &= \mathbf{J}z.
\end{align}

From \cref{eq:implicit-midpoint}, the implicit midpoint integrator computes the update,
\begin{align}
    & z' = z + \epsilon \mathbf{J}\paren{\frac{z' + z}{2}} \\
    \implies & z' = z + \frac{\epsilon}{2}\mathbf{J}z' + \frac{\epsilon}{2}\mathbf{J}z \\
    \implies& \paren{\mathrm{Id} - \frac{\epsilon}{2}\mathbf{J}}z' = \paren{\mathrm{Id} + \frac{\epsilon}{2}\mathbf{J}}z \\
    \implies& z' = \underbrace{\paren{\mathrm{Id} - \frac{\epsilon}{2}\mathbf{J}}^{-1}\paren{\mathrm{Id} + \frac{\epsilon}{2}\mathbf{J}}}_{\mathbf{Q}}z
\end{align}
The quantity $\mathbf{Q}$ is the Cayley transform of the linear transformation $\frac{\epsilon}{2}\mathbf{J}$. Noting that $\mathbf{J}$ is a skew-symmetric matrix, it is an established fact that the Cayley transform of a skew-symmetric matrix is an orthogonal matrix. This establishes that all of the eigenvalues of $\mathbf{Q}$, the linear transform representing the implicit midpoint integrator, have unit modulus.
\newpage
\section{Riemannian Metrics and a Silent Change-of-Variables?}

The Riemannian volume measure is $\sqrt{\mathrm{det}(\mathbb{G}(q))}~\mathrm{d}q$ on $\R^m$. When using HMC, does the fact that we have introduced a metric mean that we require a Jacobian correction to the posterior? Actually, the answer is no. The reason is that the Metropolis-Hastings accept-reject rule determines which density (specified with respect to the Lebesgue measure in the $(q, p)$ phase-space) is sampled by the Markov chain. Just because the acceptance Hamiltonian (which is also the guidance Hamiltonian in RMHMC; see \citet{Duane1987216}) involves computing the Riemannian metric does not mean that we have silently changed the underlying measure. Indeed, the log-determinant term appearing in \cref{eq:hamiltonian-energy} is chosen so that the conditional distribution of $p$ given $q$ is multivariate normal {\it with respect to the Lebesgue measure}.
\newpage
\section{Numerical Stability}\label{app:numerical-stability}

The stability of numerical integrators is defined by their long-term behavior on the harmonic oscillator, which is described by the following Hamiltonian system
\begin{align}
    \dot{q} = p ~~~~~~~~~~~~~~~ \dot{p} = -\omega^2 q,
\end{align}
corresponding to the Hamiltonian $H(q, p) = \omega^2 \frac{q^2}{2} + \frac{p^2}{2}$ where $\omega^2$ is a constant. Consider a single step of a numerical integrator for the harmonic oscillator with step-size $\epsilon$ that maps $(q, p)\mapsto (q', p')$. Because the harmonic oscillator is a {\it linear} differential equation, it is often possible find a matrix $\mathbf{R}\in\R^{2\times 2}$ such that $(q',p')^\top = \mathbf{R}(q,p)^\top$. A numerical method is called {\it stable} if the eigenvalues of $\mathbf{R}$ lie on the unit disk of the complex plane and are not repeated \cite{leimkuhler_reich_2005}. We have the following result.
\begin{proposition}\label{prop:implicit-midpoint-eigenvalues}
When $\epsilon < 2 / \omega$, the (generalized) leapfrog integrator is stable. The implicit midpoint integrator is stable for any $\epsilon$.
\end{proposition}
See \citet{leimkuhler_reich_2005,Hairer:1250576} for an introduction to stability analysis of numerical integrators.
\newpage
\section{Quadratic Hamiltonian}\label{app:experiment-quadratic-hamiltonian}

\begin{figure*}[b!]
    \centering
    \begin{subfigure}[b]{0.3\textwidth}
        \centering
        \includegraphics[width=\textwidth]{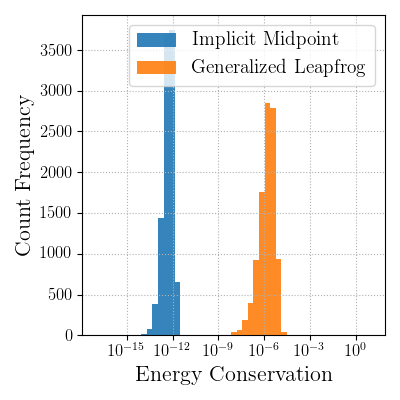}
        \caption{$\epsilon=1/100$}
    \end{subfigure}
    ~
    \begin{subfigure}[b]{0.3\textwidth}
        \centering
        \includegraphics[width=\textwidth]{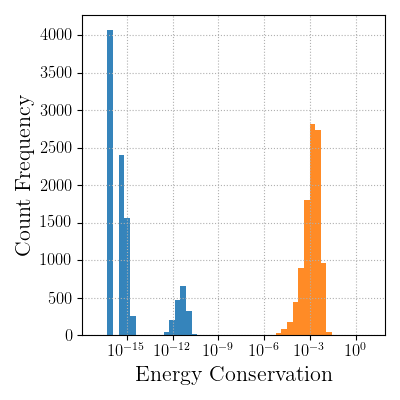}
        \caption{$\epsilon=1/10$}
    \end{subfigure}
    ~
    \begin{subfigure}[b]{0.3\textwidth}
        \centering
        \includegraphics[width=\textwidth]{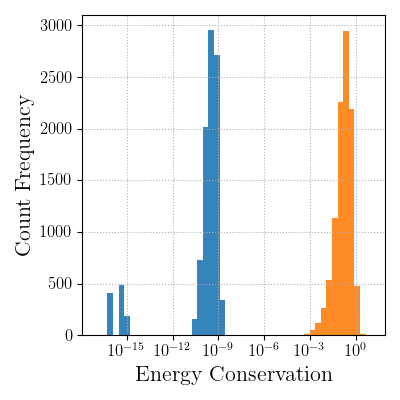}
        \caption{$\epsilon=1$}
    \end{subfigure}
    \caption{Comparison between the energy conservation of the implicit midpoint integrator and the generalized leapfrog integrator on a quadratic Hamiltonian. We observe that for every step-size, the energy conservation of the implicit midpoint method is in the neighborhood of $1\times 10^{-10}$ whereas the generalized leapfrog has energy conservation that degrades with larger steps.}
    \label{fig:gaussian-energy}
\end{figure*}

We consider using HMC to draw samples from a Gaussian distribution in two dimensions. In particular, we aim to sample from the joint distribution of position and momentum defined by,
\begin{align}
    q&\sim \mathrm{Normal}\paren{\begin{pmatrix} 1/2 \\ -1 \end{pmatrix}, \begin{pmatrix} 1 & 1/2 \\ 1/2 & 2 \end{pmatrix}} \\
    p&\sim \mathrm{Normal}\paren{\begin{pmatrix} 0 \\ 0 \end{pmatrix}, \begin{pmatrix} 1 & 1/2 \\ 1/2 & 2 \end{pmatrix}^{-1}}
\end{align}
These distributions correspond to the quadratic Hamiltonian $H(q, p)=\frac{1}{2} q^\top \Sigma^{-1} q + \frac{1}{2} p^\top\Sigma p$. This Hamiltonian can be interpreted in the Riemannian manifold setting as sampling from the posterior $q\sim\mathrm{Normal}(\mu,\Sigma)$ and the constant metric $\mathbb{G}(q) = \Sigma^{-1}$.
The corresponding Hamiltonian is quadratic and therefore \cref{thm:first-integral-implicit-midpoint} applies. We expect perfect conservation of the Hamiltonian regardless of step-size. To evaluate the conservation of the Hamiltonian energy, we consider drawing $(q, p)$ from their joint distribution and integrating Hamilton's equations of motion for ten integration steps. We consider integration step-sizes in $\set{0.01, 0.1, 1.0}$. We then compare the initial Hamiltonian energy to the Hamiltonian energy at the terminal point of the integrator. We repeat this procedure 10,000 times and show the results in \cref{fig:gaussian-energy}, where the absolute difference in Hamiltonian energy is shown as a histogram. This experiment clearly shows that the implicit midpoint integrator has excellent conservation of the quadratic Hamiltonian energy and is orders of magnitude better than the generalized leapfrog integrator. Note that for a separable Hamiltonian, as is the case here, the steps of the generalized leapfrog integrator reduce to the standard leapfrog method.
\newpage
\section{Bayesian Logistic Regression}\label{app:experiment-bayesian-logistic-regression}

Binary classification is a uniquitous task in the data sciences and logistic regression is the most popular algorithm for obtaining probabilistic estimates of class membership. Bayesian logistic regression simply equips each of the linear coefficients in the logistic regression model with a prior distribution. We consider Bayesian logistic regression as defined by the following generative model:
\begin{align}
    y_i \vert x_i,\beta  &\sim\mathrm{Bernoulli}(\sigma(x_i^\top\beta)) ~~~~~~~\mathrm{for}~ i=1,\ldots,n  \\
    \beta_i &\sim \mathrm{Normal}(0, 1) ~~~~~~~~~~~~~~~\mathrm{for}~ i=1,\ldots,k,
\end{align}
where $x_i\in \R^k$ is vector of explanatory variables and $\sigma :\R\to (0,1)$ is the sigmoid function. For the logistic regression model, let $\mathbf{x}\in\R^{n\times m}$ represent the matrix of features. The Riemannian metric formed by the sum of the Fisher information and the negative Hessian of the log-prior is $\mathbb{G}(\beta) = \mathbf{x}^\top \boldsymbol{\Lambda}\mathbf{x} + \mathrm{Id}$ where $\boldsymbol{\Lambda}$ is a diagonal matrix whose $i^\text{th}$ diagonal entry is $\sigma(x_i^\top \beta) (1-\sigma(x_i^\top \beta))$.

We consider sampling from the posterior distribution of the linear coefficients for a breast cancer, heart disease, and diabetes dataset. We consider integration step-sizes in $\set{1/10, 1}$ and a number of integration steps in $\set{5, 10, 50}$; each configuration of step-size and number of steps is replicated ten times and we attempt to draw 10,000 samples from the posterior. Results are presented in \cref{tab:logistic-regression-breast-cancer,tab:logistic-regression-diabetes,tab:logistic-regression-heart}. For the smaller step-size, both integrators enjoy very high acceptance rates and similar performance when {\it not} adjusted for timing; when adjusted for timing, the generalized leapfrog is often the better choice in the presence of a small step-size. For the larger step-size, only the implicit midpoint integrator is able to maintain a high acceptance rate; occasionally, the implicit midpoint is able to produce the optimal mean ESS and minimum ESS {\it per second}.

\clearpage
\begin{table*}[t!]
    \centering
    \scriptsize
    \begin{tabular}{lll|rrrrr}
\toprule
     &      &                            &  Acc. Prob. &   Mean ESS &    Min. ESS &  Mean ESS / Sec. &  Min. ESS / Sec. \\
Step Size & Num. Steps & Method &          &            &            &               &              \\
\midrule
0.1 & 5 & G.L.F.(a) & $1.00 \pm 0.00$ & $657.70 \pm 7.91$ & $544.12 \pm 14.20$ & $5.10 \pm 0.09$ & $4.22 \pm 0.13$ \\
& & G.L.F.(b) & $1.00 \pm 0.00$ & $656.40 \pm 8.22$ & $530.78 \pm 14.66$ & $11.64 \pm 0.19$ & $9.42 \pm 0.31$ \\
& & I.M.(a) & $1.00 \pm 0.00$ & $644.86 \pm 5.96$ & $525.95 \pm 12.06$ & $5.18 \pm 0.05$ & $4.22 \pm 0.10$ \\
& & I.M.(b) & $1.00 \pm 0.00$ & $653.46 \pm 7.75$ & $521.33 \pm 15.77$ & $5.98 \pm 0.09$ & $4.77 \pm 0.15$ \\
& 10 & G.L.F.(a) & $1.00 \pm 0.00$ & $3011.19 \pm 22.56$ & $2763.67 \pm 35.50$ & $12.33 \pm 0.10$ & $11.32 \pm 0.17$ \\
& & G.L.F.(b) & $1.00 \pm 0.00$ & $3013.59 \pm 13.75$ & $2686.52 \pm 48.17$ & $29.85 \pm 0.19$ & $26.61 \pm 0.44$ \\
& & I.M.(a) & $1.00 \pm 0.00$ & $3000.27 \pm 17.29$ & $2721.65 \pm 36.14$ & $12.60 \pm 0.09$ & $11.43 \pm 0.17$ \\
& & I.M.(b) & $1.00 \pm 0.00$ & $2982.78 \pm 15.14$ & $2716.33 \pm 43.65$ & $14.42 \pm 0.16$ & $13.13 \pm 0.27$ \\
& 50 & G.L.F.(a) & $1.00 \pm 0.00$ & $5090.37 \pm 28.81$ & $4650.52 \pm 57.69$ & $4.29 \pm 0.03$ & $3.92 \pm 0.06$ \\
& & G.L.F.(b) & $1.00 \pm 0.00$ & $5071.69 \pm 26.96$ & $4660.23 \pm 57.15$ & $11.08 \pm 0.09$ & $10.18 \pm 0.15$ \\
& & I.M.(a) & $1.00 \pm 0.00$ & $5155.52 \pm 38.72$ & $4734.41 \pm 52.70$ & $4.48 \pm 0.03$ & $4.12 \pm 0.04$ \\
& & I.M.(b) & $1.00 \pm 0.00$ & $5190.77 \pm 22.02$ & $4739.59 \pm 64.52$ & $5.23 \pm 0.05$ & $4.78 \pm 0.09$ \\ \midrule
1.0 & 5 & G.L.F.(a) & $0.19 \pm 0.02$ & $113.18 \pm 10.39$ & $61.40 \pm 9.69$ & $0.37 \pm 0.03$ & $0.20 \pm 0.03$ \\
& & G.L.F.(b) & $0.21 \pm 0.00$ & $147.14 \pm 5.94$ & $79.51 \pm 9.49$ & $1.57 \pm 0.06$ & $0.85 \pm 0.10$ \\
& & I.M.(a) & $0.88 \pm 0.00$ & $7338.66 \pm 40.87$ & $6284.89 \pm 141.01$ & $17.05 \pm 0.15$ & $14.60 \pm 0.32$ \\
& & I.M.(b) & $0.87 \pm 0.00$ & $7355.17 \pm 56.08$ & $6418.95 \pm 173.52$ & $17.93 \pm 0.16$ & $15.65 \pm 0.42$ \\
& 10 & G.L.F.(a) & $0.14 \pm 0.00$ & $181.31 \pm 22.14$ & $87.52 \pm 19.51$ & $0.45 \pm 0.05$ & $0.21 \pm 0.05$ \\
& & G.L.F.(b) & $0.14 \pm 0.00$ & $178.90 \pm 19.79$ & $90.50 \pm 17.22$ & $1.48 \pm 0.16$ & $0.75 \pm 0.14$ \\
& & I.M.(a) & $0.86 \pm 0.00$ & $31253.10 \pm 1766.12$ & $17610.80 \pm 2459.35$ & $35.43 \pm 2.34$ & $20.10 \pm 2.93$ \\
& & I.M.(b) & $0.86 \pm 0.00$ & $29505.12 \pm 1650.54$ & $17108.46 \pm 2719.39$ & $35.27 \pm 2.06$ & $20.49 \pm 3.30$ \\
& 50 & G.L.F.(a) & $0.06 \pm 0.00$ & $71.76 \pm 9.80$ & $20.62 \pm 5.41$ & $0.10 \pm 0.01$ & $0.03 \pm 0.01$ \\
& & G.L.F.(b) & $0.06 \pm 0.01$ & $108.39 \pm 25.71$ & $44.92 \pm 16.60$ & $0.47 \pm 0.10$ & $0.19 \pm 0.07$ \\
& & I.M.(a) & $0.83 \pm 0.00$ & $10127.12 \pm 327.96$ & $4430.28 \pm 278.92$ & $2.07 \pm 0.07$ & $0.91 \pm 0.06$ \\
& & I.M.(b) & $0.83 \pm 0.00$ & $10162.12 \pm 417.45$ & $4173.36 \pm 445.85$ & $2.20 \pm 0.10$ & $0.91 \pm 0.10$ \\
\bottomrule
\end{tabular}
    \caption{Comparison of the implicit midpoint and generalized leapfrog integrators on sampling from the Bayesian logistic regression posterior on the Breast Cancer dataset. For the larger step-size, the acceptance rate of the generalized leapfrog integrator completely deteriorates whereas the implicit midpoint integrator is more robust. The implicit midpoint integrator is able to achieve super-efficient sampling for a step-size of $\epsilon=1$ and ten integration steps.}
    \label{tab:logistic-regression-breast-cancer}
\end{table*}
\begin{table*}[t!]
    \centering
    \scriptsize
    \begin{tabular}{lll|rrrrr}
\toprule
     &      &                            &  Acc. Prob. &   Mean ESS &    Min. ESS &  Mean ESS / Sec. &  Min. ESS / Sec. \\
Step Size & Num. Steps & Method &          &            &            &               &              \\
\midrule
0.1 & 5 & G.L.F.(a) & $1.00 \pm 0.00$ & $651.60 \pm 10.19$ & $529.61 \pm 24.09$ & $4.21 \pm 0.06$ & $3.42 \pm 0.16$ \\
& & G.L.F.(b) & $1.00 \pm 0.00$ & $647.30 \pm 8.08$ & $544.38 \pm 14.30$ & $10.13 \pm 0.15$ & $8.52 \pm 0.23$ \\
& & I.M.(a) & $1.00 \pm 0.00$ & $649.67 \pm 5.17$ & $531.52 \pm 18.59$ & $3.84 \pm 0.06$ & $3.14 \pm 0.11$ \\
& & I.M.(b) & $1.00 \pm 0.00$ & $647.56 \pm 10.04$ & $537.62 \pm 16.70$ & $4.41 \pm 0.08$ & $3.67 \pm 0.13$ \\
& 10 & G.L.F.(a) & $1.00 \pm 0.00$ & $2986.69 \pm 23.91$ & $2705.46 \pm 58.58$ & $10.16 \pm 0.09$ & $9.20 \pm 0.19$ \\
& & G.L.F.(b) & $1.00 \pm 0.00$ & $2964.67 \pm 17.29$ & $2715.30 \pm 34.13$ & $26.06 \pm 0.39$ & $23.88 \pm 0.51$ \\
& & I.M.(a) & $1.00 \pm 0.00$ & $2959.34 \pm 21.18$ & $2741.55 \pm 35.33$ & $9.18 \pm 0.10$ & $8.51 \pm 0.12$ \\
& & I.M.(b) & $1.00 \pm 0.00$ & $2963.56 \pm 17.48$ & $2710.39 \pm 58.89$ & $10.78 \pm 0.08$ & $9.86 \pm 0.21$ \\
& 50 & G.L.F.(a) & $1.00 \pm 0.00$ & $5393.71 \pm 31.87$ & $5047.21 \pm 52.74$ & $3.80 \pm 0.03$ & $3.55 \pm 0.04$ \\
& & G.L.F.(b) & $1.00 \pm 0.00$ & $5403.67 \pm 40.17$ & $4966.93 \pm 71.68$ & $10.53 \pm 0.09$ & $9.68 \pm 0.14$ \\
& & I.M.(a) & $1.00 \pm 0.00$ & $5445.70 \pm 35.37$ & $5069.47 \pm 87.22$ & $3.51 \pm 0.03$ & $3.27 \pm 0.06$ \\
& & I.M.(b) & $1.00 \pm 0.00$ & $5529.36 \pm 24.08$ & $5161.12 \pm 39.27$ & $4.14 \pm 0.02$ & $3.86 \pm 0.03$ \\ \midrule
1.0 & 5 & G.L.F.(a) & $0.71 \pm 0.00$ & $1507.07 \pm 28.17$ & $1051.41 \pm 56.24$ & $4.78 \pm 0.10$ & $3.34 \pm 0.18$ \\
& & G.L.F.(b) & $0.71 \pm 0.00$ & $1533.13 \pm 19.75$ & $1136.48 \pm 40.48$ & $16.78 \pm 0.17$ & $12.45 \pm 0.46$ \\
& & I.M.(a) & $0.97 \pm 0.00$ & $10484.51 \pm 46.67$ & $9934.74 \pm 60.88$ & $18.74 \pm 0.08$ & $17.76 \pm 0.12$ \\
& & I.M.(b) & $0.97 \pm 0.00$ & $10446.28 \pm 28.06$ & $9870.19 \pm 65.18$ & $19.71 \pm 0.13$ & $18.62 \pm 0.17$ \\
& 10 & G.L.F.(a) & $0.60 \pm 0.00$ & $4878.12 \pm 107.69$ & $2912.57 \pm 257.17$ & $7.89 \pm 0.18$ & $4.71 \pm 0.42$ \\
& & G.L.F.(b) & $0.60 \pm 0.00$ & $5088.12 \pm 42.63$ & $3495.29 \pm 126.45$ & $30.17 \pm 0.28$ & $20.71 \pm 0.73$ \\
& & I.M.(a) & $0.97 \pm 0.00$ & $40000.00 \pm 0.00$ & $40000.00 \pm 0.00$ & $36.32 \pm 0.13$ & $36.32 \pm 0.13$ \\
& & I.M.(b) & $0.96 \pm 0.00$ & $40000.00 \pm 0.00$ & $40000.00 \pm 0.00$ & $38.39 \pm 0.08$ & $38.39 \pm 0.08$ \\
& 50 & G.L.F.(a) & $0.64 \pm 0.00$ & $6454.94 \pm 278.39$ & $4318.50 \pm 306.57$ & $2.32 \pm 0.10$ & $1.55 \pm 0.11$ \\
& & G.L.F.(b) & $0.64 \pm 0.00$ & $6357.53 \pm 179.10$ & $4245.83 \pm 213.30$ & $8.75 \pm 0.24$ & $5.84 \pm 0.29$ \\
& & I.M.(a) & $0.97 \pm 0.00$ & $37135.14 \pm 210.27$ & $20941.08 \pm 409.25$ & $6.77 \pm 0.04$ & $3.82 \pm 0.07$ \\
& & I.M.(b) & $0.97 \pm 0.00$ & $37049.46 \pm 189.78$ & $20932.03 \pm 441.83$ & $7.02 \pm 0.09$ & $3.97 \pm 0.10$ \\
\bottomrule
\end{tabular}
    \caption{Comparison of the implicit midpoint and generalized leapfrog integrators on sampling from the Bayesian logistic regression posterior on the Diabetes dataset.}
    \label{tab:logistic-regression-diabetes}
\end{table*}
\begin{table*}[t!]
    \centering
    \scriptsize
    \begin{tabular}{lll|rrrrr}
\toprule
     &      &                            &  Acc. Prob. &   Mean ESS &    Min. ESS &  Mean ESS / Sec. &  Min. ESS / Sec. \\
Step Size & Num. Steps & Method &          &            &            &               &              \\
\midrule
0.1 & 5 & G.L.F.(a) & $1.00 \pm 0.00$ & $653.84 \pm 8.93$ & $541.18 \pm 16.32$ & $3.86 \pm 0.05$ & $3.19 \pm 0.10$ \\
& & G.L.F.(b) & $1.00 \pm 0.00$ & $656.59 \pm 6.98$ & $533.13 \pm 11.07$ & $9.55 \pm 0.15$ & $7.75 \pm 0.18$ \\
& & I.M.(a) & $1.00 \pm 0.00$ & $668.14 \pm 4.07$ & $564.02 \pm 8.36$ & $4.23 \pm 0.04$ & $3.57 \pm 0.05$ \\
& & I.M.(b) & $1.00 \pm 0.00$ & $664.05 \pm 6.83$ & $517.75 \pm 15.55$ & $4.78 \pm 0.08$ & $3.73 \pm 0.14$ \\
& 10 & G.L.F.(a) & $0.99 \pm 0.00$ & $3019.18 \pm 16.31$ & $2659.45 \pm 40.27$ & $9.13 \pm 0.17$ & $8.05 \pm 0.21$ \\
& & G.L.F.(b) & $0.99 \pm 0.00$ & $3021.32 \pm 18.34$ & $2686.55 \pm 45.81$ & $24.54 \pm 0.28$ & $21.83 \pm 0.47$ \\
& & I.M.(a) & $1.00 \pm 0.00$ & $3019.84 \pm 16.66$ & $2728.01 \pm 27.65$ & $9.82 \pm 0.07$ & $8.86 \pm 0.06$ \\
& & I.M.(b) & $1.00 \pm 0.00$ & $3029.96 \pm 13.91$ & $2683.39 \pm 43.94$ & $11.42 \pm 0.10$ & $10.12 \pm 0.19$ \\
& 50 & G.L.F.(a) & $0.99 \pm 0.00$ & $4934.18 \pm 14.56$ & $4500.43 \pm 42.68$ & $3.14 \pm 0.02$ & $2.86 \pm 0.03$ \\
& & G.L.F.(b) & $0.99 \pm 0.00$ & $4952.56 \pm 14.79$ & $4431.26 \pm 54.96$ & $8.84 \pm 0.07$ & $7.91 \pm 0.11$ \\
& & I.M.(a) & $1.00 \pm 0.00$ & $5107.34 \pm 20.29$ & $4626.57 \pm 72.19$ & $3.44 \pm 0.01$ & $3.12 \pm 0.05$ \\
& & I.M.(b) & $1.00 \pm 0.00$ & $5089.77 \pm 21.30$ & $4642.54 \pm 54.89$ & $4.02 \pm 0.02$ & $3.67 \pm 0.04$ \\ \midrule
1.0 & 5 & G.L.F.(a) & $0.05 \pm 0.00$ & $25.97 \pm 3.33$ & $7.21 \pm 1.36$ & $0.08 \pm 0.01$ & $0.02 \pm 0.00$ \\
& & G.L.F.(b) & $0.05 \pm 0.00$ & $29.48 \pm 2.90$ & $9.96 \pm 2.03$ & $0.37 \pm 0.04$ & $0.13 \pm 0.03$ \\
& & I.M.(a) & $0.70 \pm 0.00$ & $4335.06 \pm 63.32$ & $3250.56 \pm 138.84$ & $6.53 \pm 0.12$ & $4.90 \pm 0.21$ \\
& & I.M.(b) & $0.70 \pm 0.00$ & $4352.91 \pm 83.82$ & $3300.84 \pm 186.04$ & $6.51 \pm 0.19$ & $4.95 \pm 0.31$ \\
& 10 & G.L.F.(a) & $0.03 \pm 0.00$ & $44.35 \pm 4.24$ & $12.20 \pm 1.90$ & $0.12 \pm 0.01$ & $0.03 \pm 0.00$ \\
& & G.L.F.(b) & $0.02 \pm 0.00$ & $44.07 \pm 8.26$ & $10.00 \pm 1.91$ & $0.51 \pm 0.11$ & $0.11 \pm 0.02$ \\
& & I.M.(a) & $0.66 \pm 0.00$ & $9217.63 \pm 305.79$ & $5320.85 \pm 661.55$ & $5.45 \pm 0.21$ & $3.16 \pm 0.41$ \\
& & I.M.(b) & $0.66 \pm 0.00$ & $9752.37 \pm 315.58$ & $5172.40 \pm 382.12$ & $5.72 \pm 0.20$ & $3.05 \pm 0.24$ \\
& 50 & G.L.F.(a) & $0.01 \pm 0.00$ & $19.19 \pm 3.58$ & $4.62 \pm 0.55$ & $0.05 \pm 0.02$ & $0.01 \pm 0.00$ \\
& & G.L.F.(b) & $0.01 \pm 0.00$ & $24.34 \pm 7.90$ & $5.81 \pm 1.41$ & $0.21 \pm 0.07$ & $0.05 \pm 0.01$ \\
& & I.M.(a) & $0.58 \pm 0.00$ & $2818.14 \pm 244.09$ & $1287.74 \pm 212.68$ & $0.22 \pm 0.02$ & $0.10 \pm 0.02$ \\
& & I.M.(b) & $0.58 \pm 0.01$ & $3028.35 \pm 370.09$ & $1445.54 \pm 329.95$ & $0.25 \pm 0.03$ & $0.12 \pm 0.03$ \\
\bottomrule
\end{tabular}
    \caption{Comparison of the implicit midpoint and generalized leapfrog integrators on sampling from the Bayesian logistic regression posterior on the heart disease dataset.}
    \label{tab:logistic-regression-heart}
\end{table*}
\clearpage
\newpage
\section{Volume Preservation and Symmetry Metrics}\label{app:volume-preservation-and-symmetric-metrics}

Here we describe how we compute metrics related to volume preservation and symmetry. Let $(q_1,\ldots, q_n)$ be samples generated by Hamiltonian Monte Carlo with numerical integrator $\Phi :\R^m\times\R^m\to\R^m\times\R^m$; each $q_i$ is an element of $\R^m$. 

\subsection{Reversibility}

For each sample $q_i$, generate $p_i\vert q_i \sim \mathrm{Normal}(0,\mathbb{G}(q_i))$ and compute $(q_i', p_i') = \Phi(q_i, p_i)$. Now compute $(q_i'', -p_i'') = \Phi(q_i', -p_i')$. The {\it violation of reversibility} is defined by
\begin{align}
    \sqrt{\Vert (q_i, p_i') - (q_i'', p_i'')\Vert_2^2}.
\end{align}
If the numerical integrator is reversible, this norm will be zero. In our metrics, we report the median violation of reversibility.

\subsection{Volume Preservation}

Let $z=(q, p)$ and identify $\Phi(z)\equiv \Phi(q, p)$. Define $f_j(z) = \frac{\Phi(z_1,\ldots, z_j + \eta / 2, \ldots, z_{2m}) - \Phi(z_1,\ldots, z_j - \eta / 2, \ldots, z_{2m})}{\eta}$ for $\eta = 1\times 10^{-5}$ (except for the Fitzhugh-Nagumo model where we set $\eta=1\times 10^{-3}$ for numerical reasons), which is the central difference formula that approximates $\frac{\partial}{\partial z_j} \Phi(z)$. We compute the approximation to the Jacobian of $\Phi$ by constructing,
\begin{align}
    \nabla \Phi(z) \approx F(z) \defeq \begin{pmatrix} f_1(z) & f_2(z) &\cdots & f_{2m}(z) \end{pmatrix} \in\R^{2m\times2m}.
\end{align}
For each sample $q_i$, generate $p_i\vert q_i \sim \mathrm{Normal}(0,\mathbb{G}(q_i))$ and set $z_i = (q_i, p_i)$. 
The {\it violation of volume preservation} is defined by
\begin{align}
    \left|\mathrm{det}(F(z_i)) - 1\right|.
\end{align}
If the numerical integrator is volume preserving, this difference will be zero.  In our metrics, we report the median violation of volume preservation.
\newpage
\section{Detailed Balance from Reversibility and Volume Preservation}\label{app:detailed-balance}

Let $\Phi : \R^{m}\times\R^m\to\R^m\times \R^m$ be a numerical integrator satisfying the following two properties as described in the main text:
\begin{enumerate}[(i)]
    \item The integrator has a unit Jacobian determinant so that it preserves volume in $(q, p)$-space.
    \item The integrator is symmetric under negation of the momentum variable.
\end{enumerate}

Given $(q, p)\in\R^m\times\R^m$, define the {\it momentum flip operator} by $\Psi(q, p) = (q, -p)$. For Markov chain Monte Carlo, we then define the Markov chain {\it proposal} operator by $\mathfrak{P} \defeq \Psi\circ\Phi$; because $\Phi$ satisfies property (ii) we have that $\mathfrak{P}\circ \mathfrak{P} = \Psi\circ\Phi \circ \Psi\circ\Phi = \mathrm{Id}$ so that the proposal operator is self-inverse $\mathfrak{P}^{-1} = \mathfrak{P}$. Moreover, $\Psi$ has unit Jacobian determinant: $\left|\mathrm{det}(\nabla\Psi)\right| = 1$; therefore, since $\Phi$ has unit Jacobian determinant by property (i), $\mathfrak{P}$ also has unit Jacobian determinant:
\begin{align}
    \left|\mathrm{det}(\nabla\mathfrak{P}(q, p)) \right| &= \left|\mathrm{det}(\nabla \Psi (\Phi(q, p)) \nabla \Phi(q, p))\right| \\
    &= \left|\mathrm{det}(\nabla \Psi (\Phi(q, p)) \cdot \mathrm{det}(\nabla \Phi(q, p))\right| \\
    &= 1
\end{align}

Given a current position $(q, p)$ of the Markov chain, the proposal for the next state of the Markov chain $(q',p') \defeq \mathfrak{P}(q, p)$ is accepted with probability
\begin{align}
    \min\set{1, \frac{\pi(q',p')}{\pi(q, p)}},
\end{align}
where $\pi:\R^m\times\R^m\to\R$ is the target distribution. (Notice that $\pi$ must only be specified up to a constant.) Thus, the Markov chain {\it transition} operator is defined by,
\begin{align}
    \mathfrak{T}(q, p) = \begin{cases}
    (q', p') & \mathrm{w.p.}~ \min\set{1, \frac{\pi(q',p')}{\pi(q, p)}} \\
    (q, p) & \mathrm{else}.
    \end{cases}
\end{align}

We say that detailed balance holds if for all sets $A, B\subset\R^m\times\R^m$ we have,
\begin{align}
    \underset{(q, p)\sim \pi}{\mathrm{Pr}}\left[(q,p)\in A ~\mathrm{and}~ \mathfrak{T}(q, p)\in B\right] = \underset{(q, p)\sim \pi}{\mathrm{Pr}}\left[(q,p)\in B ~\mathrm{and}~ \mathfrak{T}(q, p)\in A\right]
\end{align}
Let $z=(q, p)$. Expanding we compute,
\begin{align}
    \underset{z\sim \pi}{\mathrm{Pr}}\left[z\in A ~\mathrm{and}~ \mathfrak{T}(z)\in B\right] &= \int_A \pi(z) \cdot \mathbf{1}\set{\mathfrak{P}(z) \in B}\cdot \min\set{1, \frac{\pi(\mathfrak{P}(z))}{\pi(z)}}~\mathrm{d}z \\
    &= \int_{A\cap \mathfrak{P}(B)} \pi(z) \cdot \min\set{1, \frac{\pi(\mathfrak{P}(z))}{\pi(z)}}~\mathrm{d}z \\
    &\overset{z' = \mathfrak{P}(z)}{=} \int_{\mathfrak{P}(A)\cap B} \pi(\mathfrak{P}(z')) \cdot \min\set{1, \frac{\pi(z')}{\pi(\mathfrak{P}(z'))}}~\mathrm{d}z' \label{eq:change-of-variables} \\
    &= \int_{\mathfrak{P}(A)\cap B} \pi(z')  \cdot \min\set{1, \frac{\pi(\mathfrak{P}(z'))}{\pi(z')}}~\mathrm{d}z' \\
    &= \int_{B} \pi(z') \cdot \mathbf{1}\set{\mathfrak{P}(z')\in A}  \cdot \min\set{1, \frac{\pi(\mathfrak{P}(z'))}{\pi(z')}}~\mathrm{d}z' \\
    &= \underset{z\sim \pi}{\mathrm{Pr}}\left[z\in B ~\mathrm{and}~ \mathfrak{T}(z)\in A\right]
\end{align}
showing that detailed balance holds. The change-of-variables in \cref{eq:change-of-variables} does not incur a Jacobian determinant correction since $\mathfrak{P}$ has unit Jacobian determinant.

\newpage
\section{Implementation of Integrators}\label{app:implementation-of-integrators}

\subsection{Implementations of the Generalized Leapfrog Integrator}

As the purpose of this research is to compare two integrators for Hamiltonian Monte Carlo, we wish to be precise about how these numerical methods have been implemented. 

\begin{algorithm}[b!]
\caption{{\bf (G.L.F.(b))} The procedure for a single step of integrating Hamiltonian dynamics using the efficient implementation of the generalized leapfrog integrator.}
\label{alg:smart-generalized-leapfrog}
\begin{algorithmic}[1]
\STATE \textbf{Input}: Log-posterior $\mathcal{L}:\R^m\to \R$, Riemannian metric $\mathbb{G}:\R^m\to\R^{m\times m}$, initial position and momentum variables $(q,p)\in\R^m\times\R^m$, integration step-size size $\epsilon\in\R$.
\STATE Precompute $\mathbb{G}^{-1}(q)$, $\frac{\partial}{\partial q_i}\mathbb{G}(q)$, and define
\begin{align}
    A_i \defeq \frac{\partial}{\partial q_i} \mathcal{L}(q) + \frac{1}{2}\mathrm{trace}\paren{\mathbb{G}^{-1}(q) \frac{\partial}{\partial q_i}\mathbb{G}(q)}
\end{align}
for $i=1,\ldots, m$.
\STATE Use \cref{alg:fixed-point-solver} with tolerance $\delta$ and initial guess $p$ to solve for $\bar{p}$,
\begin{align}
    \label{eq:smart-generalized-leapfrog-momentum} \bar{p} &= \underbrace{p - \frac{\epsilon}{2}\paren{A_1 - \frac{1}{2} (\mathbb{G}^{-1}(q) \bar{p})^\top \frac{\partial}{\partial q_1}\mathbb{G}(q) (\mathbb{G}^{-1}(q) \bar{p}), \ldots, A_m - \frac{1}{2} (\mathbb{G}^{-1}(q) \bar{p})^\top \frac{\partial}{\partial q_m}\mathbb{G}(q) (\mathbb{G}^{-1}(q) \bar{p})}^\top}_{f(\bar{p})}.
\end{align}
\STATE Precompute $\mathbb{G}(q)^{-1}\bar{p}$.
\STATE Use \cref{alg:fixed-point-solver} with tolerance $\delta$ and initial guess $q$ to solve for $q'$,
\begin{align}
    \label{eq:smart-generalized-leapfrog-position} q' = \underbrace{q + \frac{\epsilon}{2}\paren{\mathbb{G}^{-1}(q)\bar{p} + \mathbb{G}^{-1}(q')\bar{p}}}_{f(q')}.
\end{align}
\STATE Compute $p'$ using \cref{eq:glf-momentum-ii}, which is an explicit update.
\STATE \textbf{Return}: $(q', p')\in\R^m\times\R^m$.
\end{algorithmic}
\end{algorithm}

The first implementation of the generalized leapfrog integrator is presented in \cref{alg:generalized-leapfrog}. 
The system \cref{eq:glf-momentum-i,eq:glf-position,eq:glf-momentum-ii} was described as ``naive'' because it appears to ignore important structural properties of the equations of motion in \cref{eq:position-evolution,eq:momentum-evolution} that would accelerate a step of the generalized leapfrog integrator. For instance, in \cref{eq:glf-momentum-i}, the metric $\mathbb{G}(q)$ may be precomputed because it is an invariant of the fixed-point relation. Another example is that $\mathbb{G}^{-1}(q)\bar{p}$ is an invariant quantity of \cref{eq:glf-position} and need not be recomputed in each fixed-point iteration. Thus, we see that the generalized leapfrog integrator, when efficiently implemented, has an important computational advantage in that certain invariant quantities can be ``cached'' when finding fixed points. A more complicated implementation of the generalized leapfrog integrator with caching is presented in given in \cref{alg:smart-generalized-leapfrog}. We stress that \cref{alg:generalized-leapfrog,alg:smart-generalized-leapfrog} perform the same calculation.

\subsection{Implementations of the Implicit Midpoint Integrator}

\begin{algorithm}[t!]
\caption{{\bf (I.M.(b))} The procedure for a single step of integrating Hamiltonian dynamics using the implicit midpoint integrator as advocated by \citet{leimkuhler_reich_2005}.}
\label{alg:smart-implicit-midpoint}
\begin{algorithmic}[1]
\STATE \textbf{Input}: Hamiltonian $H:\R^m\times\R^m\to \R$, initial position and momentum variables $(q,p)\in\R^m\times\R^m$, integration step-size size $\epsilon\in\R$, fixed-point convergence tolerance $\delta\geq 0$.
\STATE Use \cref{alg:fixed-point-solver} with tolerance $\delta$ and initial guess $(q, p)$ to solve for $(\bar{q},\bar{p})$, \begin{align}
    \label{eq:implicit-midpoint-implicit} \begin{pmatrix} \bar{q} \\ \bar{p} \end{pmatrix} \defeq \underbrace{\begin{pmatrix} q \\ p \end{pmatrix} + \frac{\epsilon}{2} \begin{pmatrix} \nabla_p H(\bar{q}, \bar{p}) \\ -\nabla_q H(\bar{q},\bar{p}) \end{pmatrix}}_{f(\bar{q},\bar{p})}.
\end{align}
\STATE Compute the explicit update
\begin{align}
    \label{eq:implicit-midpoint-explicit} \begin{pmatrix} q' \\ p' \end{pmatrix} =  \begin{pmatrix} \bar{q} \\ \bar{p} \end{pmatrix} + \frac{\epsilon}{2} \begin{pmatrix} \nabla_p H(\bar{q}, \bar{p}) \\ -\nabla_q H(\bar{q},\bar{p}) \end{pmatrix}.
\end{align}
\STATE \textbf{Return}: $(q', p')\in\R^m\times\R^m$.
\end{algorithmic}
\end{algorithm}

In addition to the implementation of the implicit midpoint method described in \cref{alg:implicit-midpoint}, we also consider a variant advocated by \citet{leimkuhler_reich_2005} and present the implementation in \cref{alg:smart-implicit-midpoint}. The essential difference between \cref{alg:implicit-midpoint,alg:smart-implicit-midpoint} is whether or not the implicitly-defined update computes the terminal point of the step (\cref{alg:implicit-midpoint}) or the midpoint of the step (\cref{alg:smart-implicit-midpoint}). We stress that \cref{alg:implicit-midpoint,alg:smart-implicit-midpoint} perform the same calculation when $\delta=0$, which is easily verified by plugging \cref{eq:implicit-midpoint-implicit} into \cref{eq:implicit-midpoint-explicit} and comparing to \cref{eq:implicit-midpoint}. Unlike the generalized leapfrog integrator, the implicit midpoint integrator does not enjoy the ability to cache intermediate computations.

\subsection{Remarks on Computational Complexity}\label{app:lipschitz-fixed-point}

In discussing the computational complexity of the generalized leapfrog and implicit midpoint methods, we will assume that the functions defining the fixed-point relations are {\it contraction maps}. Contraction maps in a complete metric space guarantee that fixed point equations have unique solutions and that these solutions are reached via fixed point iteration for any initial condition. Moreover, contraction maps have known convergence rates, which are convenient for analysis.

\begin{definition}
The $L_\infty$ distance on $\R^m$ is defined by
\begin{align}
    d_\infty(z, z') \defeq \max_{i=1,\ldots, m} \abs{z_i - z'_i}.
\end{align}
\end{definition}
We will restrict our attention to the $L_\infty$ distance as it is used as our convergence criterion in \cref{alg:fixed-point-solver}.
\begin{definition}
A contraction map on $\R^m$ is a Lipschitz function $f:\R^m\to\R^m$ whose Lipschitz constant is less than one; that is,
\begin{align}
   d_\infty(f(z), f(z')) \leq L \cdot d_\infty(z, z')
\end{align}
for all $z, z'\in\R^m$ and where $L\in [0, 1)$.
\end{definition}

\begin{theorem}[Banach Fixed Point Theorem]
Let $f :\R^m\to\R^m$ be a contraction map with Lipschitz constant $L\in [0, 1)$. Then (i) the equation $z=f(z)$ has a unique solution; (ii) the fixed point iterations $z_{n+1} = f(z_n)$ converge to $z$ from any $z_0\in\R^m$; (iii) the distance between iterates satisfies $d_\infty(z_{n+1},z_n) \leq L^n d_\infty(z_1, z_0)$.
\end{theorem}

In assessing convergence to the unique solution using fixed point iteration, \cref{alg:fixed-point-solver} demands that $d_\infty(z_{n+1},z_n) \leq \delta$ for some convergence tolerance $\delta >0$. Assuming that $d_\infty(z_1, z_0) > 0$, we can rearrange $L^n d_\infty(z_1, z_0) \leq \delta$ to give a sufficient condition on the number of fixed point iterates $n$ to guarantee that $d_\infty(z_{n+1},z_n) \leq \delta$. Namely:
\begin{lemma}
Assume $d_\infty(z_1, z_0) > 0$. Then the choice
\begin{align}
    n = \ceil[\Bigg]{\frac{\log \delta - \log d_\infty(z_1, z_0)}{\log L}}
\end{align}
is sufficient to ensure that $d_\infty(z_{n+1},z_n) \leq \delta$.
\end{lemma}
\begin{proof}
From the Banach fixed point theorem we know that $d_\infty(z_{n+1},z_n) \leq L^n d_\infty(z_1, z_0)$. Therefore, we seek to establish when the right-hand side of the inequality is less than $\delta$. Rearranging and noting that $\log L$ is negative (since $f$ is a contraction map) yields
\begin{align}
    n \geq \frac{\log \delta - \log d_\infty(z_1, z_0)}{\log L}.
\end{align}
It makes sense to use the smallest integer $n$ such that the $d_\infty(z_{n+1},z_n) \leq \delta$ so computing the ceiling of the right-hand side yields the result.
\end{proof}

Under the assumption that functions defining fixed point equations are contractions, we can give an approximate comparison of the computational complexity of the implicit midpoint and generalized leapfrog algorithms. In integrating the dynamics corresponding to a Hamiltonian as in \cref{eq:hamiltonian-energy} (refer to \cref{eq:position-evolution,eq:momentum-evolution}), there are several operations of notable computational complexity; these are (i) computing the gradient of the log-likelihood, (ii) computing the inverse of the Riemannian metric, and (iii) computing the derivatives of the Riemannian metric. To make this slightly more formal, let $\mathrm{Cost}(\nabla \mathcal{L})$, $\mathrm{Cost}(\mathbb{G}^{-1})$, and $\mathrm{Cost}(\nabla \mathbb{G})$ denote some notion of computational complexity associated to these three quantities; we assume that any remaining arithmetic operations used in the integration of Hamilton's mechanics have negligible computational cost.

\begin{example}
For the implicit midpoint integrator, each of the above quantities (i)-(iii) must be computed within each fixed point iteration. If $L_\mathrm{imp}$ denotes the Lipschitz constant associated to the map $f$ in \cref{eq:implicit-midpoint}, then the total computational cost of the implicit midpoint integrator is
\begin{align}
    \label{eq:cost-implicit-midpoint} \ceil[\Bigg]{\frac{\log \delta - \log d_\infty(f(z_0), z_0)}{\log L_\mathrm{imp}}} \cdot \paren{\mathrm{Cost}(\nabla \mathcal{L}) + \mathrm{Cost}(\mathbb{G}^{-1}) + \mathrm{Cost}(\nabla \mathbb{G})}.
\end{align}
\end{example}
\begin{example}
The situation is rather different for the generalized leapfrog integrator as expressed in \cref{alg:smart-generalized-leapfrog}. The fixed point equation in \cref{eq:smart-generalized-leapfrog-momentum} requires that we compute quantities (i)-(iii); however, these are invariant of the fixed point equation in \cref{eq:smart-generalized-leapfrog-momentum}, which therefore does not incur additional cost beyond the sum of $\mathrm{Cost}(\nabla \mathcal{L})$, $\mathrm{Cost}(\mathbb{G}^{-1})$, and $\mathrm{Cost}(\nabla \mathbb{G})$. For the second fixed point equation, let $L_\mathrm{glf}$ denote the Lipschitz constant of the contraction map $f$ in \cref{eq:smart-generalized-leapfrog-position}; each iteration requires computing the inverse of the Riemannian metric. Thus, the computational cost incurred by the second fixed point equation is
\begin{align}
    \ceil[\Bigg]{\frac{\log \delta - \log d_\infty(f(q_0), q_0)}{\log L_\mathrm{glf}}} \cdot \mathrm{Cost}(\mathbb{G}^{-1}).
\end{align}
The final step of the generalized leapfrog integrator requires an explicit update wherein we compute quantities (i)-(iii) using the updated position variable. In total, the cost of a step of the generalized leapfrog integrator is therefore,
\begin{align}
    \label{eq:cost-generalized-leapfrog} \ceil[\Bigg]{\frac{\log \delta - \log d_\infty(f(q_0), q_0)}{\log L_\mathrm{glf}}} \cdot \mathrm{Cost}(\mathbb{G}^{-1}) + 2\cdot \paren{\mathrm{Cost}(\nabla \mathcal{L}) + \mathrm{Cost}(\mathbb{G}^{-1}) + \mathrm{Cost}(\nabla \mathbb{G})}.
\end{align}
\end{example}

\subsection{Counts of Fixed Point Iterations}

\begin{figure*}[t!]
    \centering
    \includegraphics[width=\textwidth]{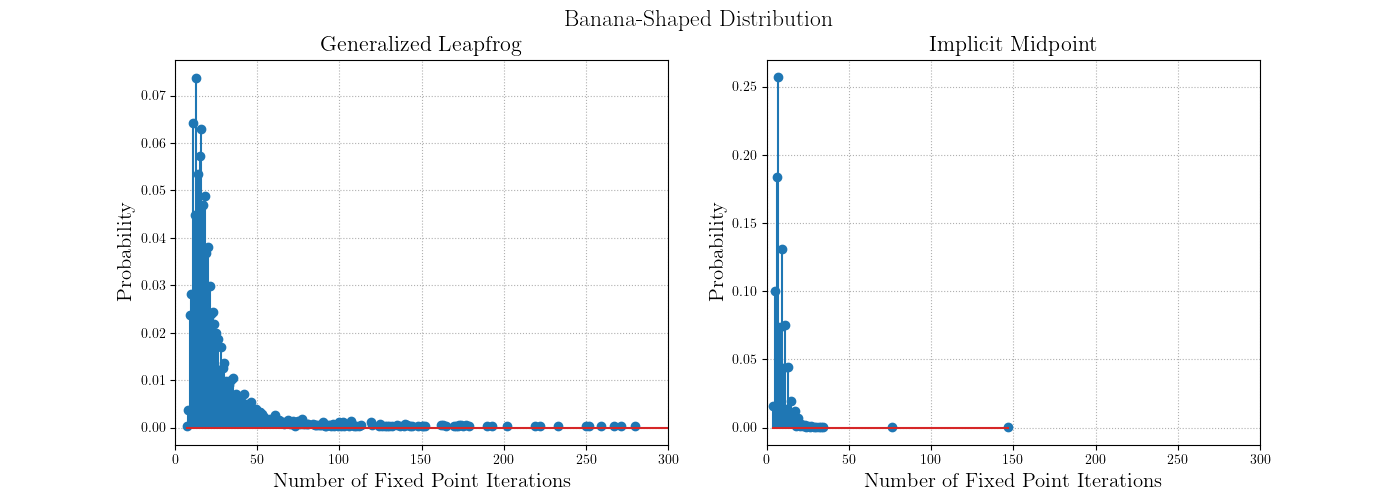}
    \caption{Visualization of the relative frequency of the number of fixed point iterations in the implicit midpoint and generalized leapfrog integration procedure for the banana-shaped distribution.}
    \label{fig:banana-fixed-point}
\end{figure*}
\begin{figure*}[t!]
    \centering
    \includegraphics[width=\textwidth]{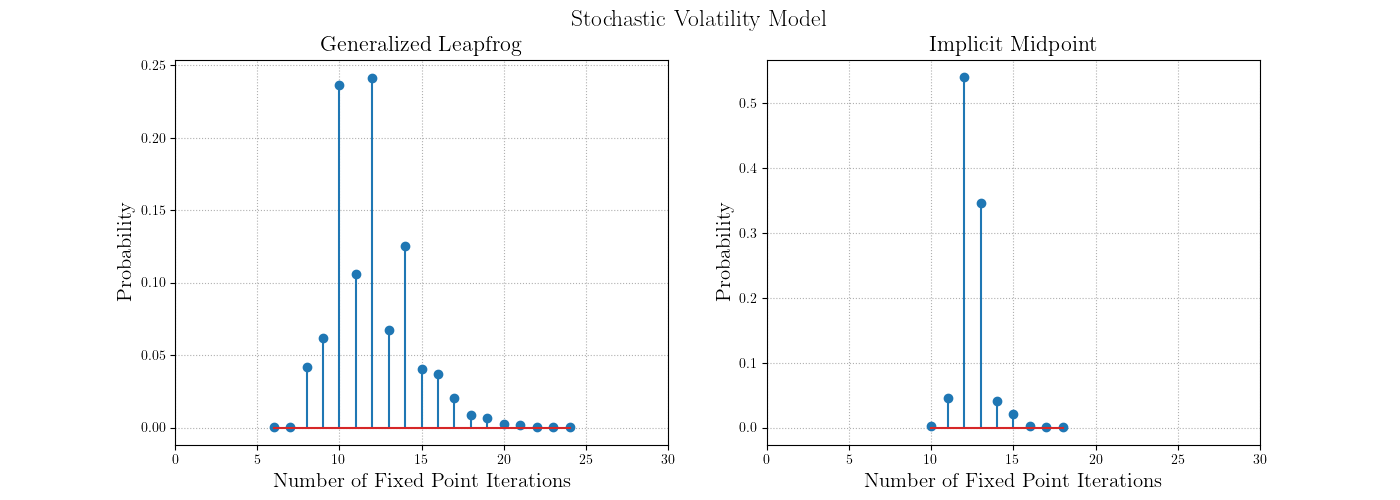}
    \caption{Visualization of the relative frequency of the number of fixed point iterations in the implicit midpoint and generalized leapfrog integration procedure for the stochastic volatility model.}
    \label{fig:stochastic-volatility-fixed-point}
\end{figure*}
\begin{figure*}[t!]
    \centering
    \includegraphics[width=\textwidth]{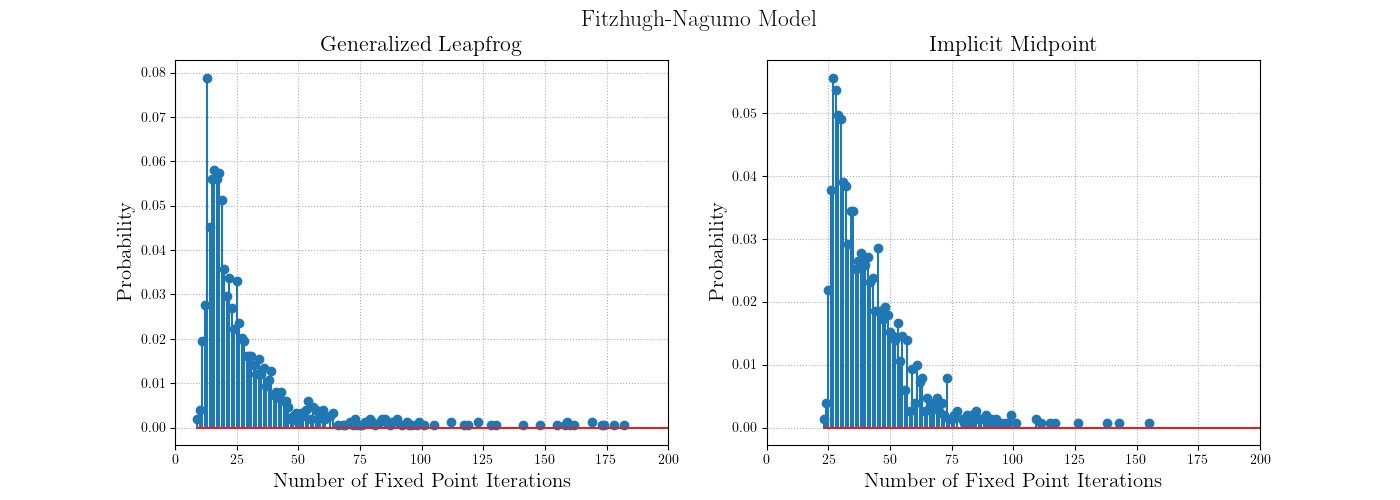}
    \caption{Visualization of the relative frequency of the number of fixed point iterations in the implicit midpoint and generalized leapfrog integration procedure for the Fitzhugh-Nagumo model.}
    \label{fig:fitzhugh-nagumo-fixed-point}
\end{figure*}

In this section we compute the number of fixed point iterations performed by both the implicit midpoint and generalized leapfrog integrators on the banana, stochastic volatility, and Fitzhugh-Nagumo experiments.  We report the relative frequency of each number of the fixed point iterations over the course of sampling. We note that this is for context only, and that the precise number of fixed point iterations consumed by either integrator are not comparable to one another; the reason for this is that the amount of computation differs for the implicit midpoint and generalized leapfrog methods as described in \cref{app:lipschitz-fixed-point}. For the generalized leapfrog method, we report the total number of iterations to solve both fixed point relations to the prescribed tolerance. Results are shown in \cref{fig:stochastic-volatility-fixed-point,fig:fitzhugh-nagumo-fixed-point,fig:banana-fixed-point} for a convergence tolerance of $\delta = 1\times 10^{-6}$.
\clearpage
\section{Proof of \Cref{prop:implicit-midpoint-quadratic-hamiltonian}}\label{app:implicit-midpoint-quadratic-hamiltonian}

\begin{proof}
The implicit midpoint integrator conserves quadratic first integrals (see \cref{thm:first-integral-implicit-midpoint}) and, in this case, the Hamiltonian energy is itself a quadratic function. 
\end{proof}
\clearpage
\section{Randomized Step Experimental Design}\label{app:randomized-step-experimental-design}

In the main text we have considered a grid search over the number of integration steps and use this number of steps in generating every proposal using either the implicit midpoint or generalized leapfrog integrators. An alternative experimental design is to randomize the number of integration steps for each proposal by sampling the number of steps uniformly between one and some upper bound. When the Hamiltonian is separable, the leapfrog integrator with a single step recovers the Metropolis-adjusted Langevin algorithm, which has known ergodicity properties. Therefore, it may be anticipated that randomizing the number of integration steps may lead to improved ergodicity when the proposal is computed by the generalized leapfrog method. In this appendix, we consider randomizing the number of integration steps for both the implicit midpoint and generalized leapfrog integrators.

\subsection{Banana-Shaped Distribution}

\begin{table*}[b!]
    \centering
    \scriptsize
    \begin{tabular}{lll|rrrrrr}
\toprule
     &      &                            &  Acc. Prob. & Time (Sec.) &   Mean ESS &    Min. ESS &  Mean ESS / Sec. &  Min. ESS / Sec. \\
Step Size & Max. Steps & Method &          &            &            &               &              \\
\midrule
0.1 & 5 & G.L.F.(a) & $0.66 \pm 0.01$ & $298.11 \pm 4.99$ & $240.50 \pm 13.01$ & $140.54 \pm 10.54$ & $0.81 \pm 0.04$ & $0.47 \pm 0.03$ \\
& & G.L.F.(b) & $0.66 \pm 0.01$ & $103.90 \pm 1.64$ & $246.99 \pm 10.57$ & $149.70 \pm 8.26$ & $2.37 \pm 0.08$ & $1.44 \pm 0.07$ \\
& & I.M.(a) & $0.99 \pm 0.00$ & $65.48 \pm 0.97$ & $372.91 \pm 20.52$ & $276.00 \pm 18.41$ & $5.69 \pm 0.28$ & $4.20 \pm 0.24$ \\
& & I.M.(b) & $0.99 \pm 0.00$ & $59.90 \pm 0.98$ & $375.99 \pm 23.34$ & $268.26 \pm 15.80$ & $6.32 \pm 0.45$ & $4.51 \pm 0.30$ \\
& 10 & G.L.F.(a) & $0.60 \pm 0.01$ & $420.38 \pm 8.68$ & $541.71 \pm 13.63$ & $373.56 \pm 16.10$ & $1.29 \pm 0.05$ & $0.89 \pm 0.04$ \\
& & G.L.F.(b) & $0.59 \pm 0.00$ & $153.94 \pm 3.34$ & $560.73 \pm 12.99$ & $370.12 \pm 13.86$ & $3.66 \pm 0.12$ & $2.41 \pm 0.11$ \\
& & I.M.(a) & $0.98 \pm 0.00$ & $111.84 \pm 1.64$ & $1205.85 \pm 38.74$ & $945.21 \pm 35.13$ & $10.80 \pm 0.37$ & $8.47 \pm 0.34$ \\
& & I.M.(b) & $0.98 \pm 0.00$ & $100.24 \pm 1.79$ & $1212.95 \pm 41.28$ & $948.95 \pm 47.69$ & $12.12 \pm 0.42$ & $9.47 \pm 0.46$ \\
& 50 & G.L.F.(a) & $0.31 \pm 0.00$ & $1158.96 \pm 13.35$ & $744.51 \pm 34.97$ & $603.15 \pm 17.87$ & $0.64 \pm 0.03$ & $0.52 \pm 0.02$ \\
& & G.L.F.(b) & $0.31 \pm 0.01$ & $451.75 \pm 4.30$ & $754.07 \pm 37.32$ & $602.83 \pm 42.83$ & $1.67 \pm 0.09$ & $1.34 \pm 0.10$ \\
& & I.M.(a) & $0.97 \pm 0.00$ & $473.24 \pm 4.62$ & $6306.89 \pm 136.15$ & $2568.88 \pm 34.70$ & $13.33 \pm 0.28$ & $5.43 \pm 0.07$ \\
& & I.M.(b) & $0.97 \pm 0.00$ & $419.67 \pm 3.65$ & $6345.83 \pm 165.83$ & $2597.81 \pm 50.49$ & $15.14 \pm 0.46$ & $6.20 \pm 0.14$ \\
\bottomrule
\end{tabular}
    \caption{Comparison of the implicit midpoint and generalized leapfrog integrators on sampling from the banana-shaped distribution when a randomized number of steps is used.}
    \label{tab:randomized-step-banana-ess}
\end{table*}

Results for the banana-shaped distribution when using the randomized number of steps experimental design are presented in \cref{tab:randomized-step-banana-ess}.

\subsection{Neal Funnel}

\begin{table*}[b!]
    \centering
    \scriptsize
    \begin{tabular}{lll|rrrrrr}
\toprule
     &      &                            &  Acc. Prob. & Time (Sec.) &   Mean ESS &    Min. ESS &  Mean ESS / Sec. &  Min. ESS / Sec. \\
Max. Steps & Step Size & Method &          &            &            &               &              \\
\midrule
20 & 0.1 & G.L.F.(a) & $0.99 \pm 0.00$ & $786.17 \pm 62.99$ & $3535.77 \pm 45.31$ & $120.34 \pm 15.07$ & $4.69 \pm 0.31$ & $0.16 \pm 0.02$ \\
& & I.M.(a) & $1.00 \pm 0.00$ & $655.70 \pm 33.04$ & $3538.75 \pm 28.40$ & $132.25 \pm 12.73$ & $5.48 \pm 0.21$ & $0.21 \pm 0.02$ \\
& & I.M.(b) & $1.00 \pm 0.00$ & $579.39 \pm 40.26$ & $3545.94 \pm 32.44$ & $133.26 \pm 10.57$ & $6.29 \pm 0.31$ & $0.23 \pm 0.02$ \\
& 0.2 & G.L.F.(a) & $0.96 \pm 0.00$ & $968.70 \pm 58.97$ & $11305.65 \pm 109.81$ & $525.33 \pm 15.22$ & $11.96 \pm 0.59$ & $0.56 \pm 0.04$ \\
& & I.M.(a) & $0.99 \pm 0.00$ & $818.02 \pm 32.74$ & $11809.38 \pm 128.38$ & $502.90 \pm 33.26$ & $14.59 \pm 0.51$ & $0.63 \pm 0.05$ \\
& & I.M.(b) & $0.99 \pm 0.00$ & $728.94 \pm 21.80$ & $11959.25 \pm 115.00$ & $522.38 \pm 22.78$ & $16.52 \pm 0.51$ & $0.72 \pm 0.04$ \\
& 0.5 & G.L.F.(a) & $0.43 \pm 0.01$ & $1420.67 \pm 49.91$ & $1857.57 \pm 172.54$ & $754.50 \pm 85.59$ & $1.30 \pm 0.11$ & $0.53 \pm 0.06$ \\
& & I.M.(a) & $0.89 \pm 0.00$ & $1596.95 \pm 48.55$ & $8468.24 \pm 54.32$ & $3018.45 \pm 93.75$ & $5.34 \pm 0.15$ & $1.91 \pm 0.11$ \\
& & I.M.(b) & $0.89 \pm 0.00$ & $1458.31 \pm 53.48$ & $8422.27 \pm 57.72$ & $3034.97 \pm 81.87$ & $5.83 \pm 0.19$ & $2.10 \pm 0.09$ \\
\bottomrule
\end{tabular}
    \caption{Comparison of the implicit midpoint and the naive generalized leapfrog integrators on sampling from Neal's funnel distribution when using a randomized number of integration steps.}
    \label{tab:randomized-step-neal-funnel-ess}
\end{table*}

Results for Neal's funnel distribution when using the randomized number of steps experimental design are presented in \cref{tab:randomized-step-neal-funnel-ess}.

\subsection{Stochastic Volatility Model}

\begin{table*}[t!]
    \centering
    \tiny
    \begin{tabular}{l|rrrrrr|rr|rr}
\toprule
       &&&&&&& \multicolumn{2}{c}{Volume Preservation} & \multicolumn{2}{c}{Symmetry} \\
    Method   & Acc. Prob. & Time (Sec.) &   Mean ESS &    Min. ESS &  \begin{tabular}[c]{@{}l@{}}Mean ESS \\ (Sec.)\end{tabular} &  \begin{tabular}[c]{@{}l@{}}Min. ESS \\ (Sec.)\end{tabular} & Median &    $90^\mathrm{th}$-Per. &  Median &    $90^\mathrm{th}$-Per. \\
\midrule
G.L.F.(a) & $0.85 \pm 0.0$ & $2367.93 \pm 16.1$ & $335.80 \pm 15.3$ & $132.96 \pm 3.9$ & $0.14 \pm 0.0$ & $0.06 \pm 0.0$ & $6.8\mathrm{e-}07$ & $2.9\mathrm{e-}06$ & $4.9\mathrm{e-}06$ & $2.4\mathrm{e-}05$ \\
G.L.F.(b) & $0.85 \pm 0.0$ & $2196.52 \pm 8.5$ & $331.84 \pm 12.1$ & $139.30 \pm 3.7$ & $0.15 \pm 0.0$ & $0.06 \pm 0.0$ & $7.1\mathrm{e-}07$ & $2.9\mathrm{e-}06$ & $4.8\mathrm{e-}06$ & $2.7\mathrm{e-}05$ \\
I.M.(a) & $0.86 \pm 0.0$ & $2434.74 \pm 7.9$ & $316.98 \pm 15.4$ & $131.79 \pm 3.4$ & $0.13 \pm 0.0$ & $0.05 \pm 0.0$ & $8.7\mathrm{e-}08$ & $2.3\mathrm{e-}07$ & $1.3\mathrm{e-}06$ & $2.4\mathrm{e-}06$ \\
I.M.(b) & $0.87 \pm 0.0$ & $2399.22 \pm 6.4$ & $342.11 \pm 14.0$ & $133.76 \pm 3.7$ & $0.14 \pm 0.0$ & $0.06 \pm 0.0$ & $1.7\mathrm{e-}07$ & $4.4\mathrm{e-}07$ & $2.8\mathrm{e-}06$ & $4.7\mathrm{e-}06$ \\
\bottomrule
\end{tabular}
    \caption{Comparison of the implicit midpoint generalized leapfrog integrators on the stochastic volatility model when using a randomized number of integration steps.}
    \label{tab:randomized-step-stochastic-volatility-symmetry-volume-preservation}
\end{table*}

Results for the stochastic volatility model when using the randomized number of steps experimental design are presented in \cref{tab:randomized-step-stochastic-volatility-symmetry-volume-preservation}.

\subsection{Fitzhugh-Nagumo Differential Equation Model}

\begin{table*}[t!]
    \centering
    \scriptsize
    \begin{tabular}{lll|rrrrrr}
\toprule
     &      &                            &  Acc. Prob. & Time (Sec.) &   Mean ESS &    Min. ESS &  Mean ESS / Sec. &  Min. ESS / Sec. \\
Step Size & Num. Steps & Method &          &            &            &               &              \\
\midrule
1.0 & 1 & G.L.F.(a) & $0.75 \pm 0.01$ & $4084.16 \pm 138.81$ & $289.87 \pm 12.89$ & $251.36 \pm 12.33$ & $0.07 \pm 0.01$ & $0.06 \pm 0.00$ \\
& & G.L.F.(b) & $0.74 \pm 0.01$ & $1193.08 \pm 61.74$ & $301.44 \pm 16.67$ & $262.36 \pm 15.38$ & $0.26 \pm 0.02$ & $0.23 \pm 0.02$ \\
& & I.M.(a) & $0.95 \pm 0.00$ & $4353.40 \pm 104.50$ & $237.73 \pm 8.67$ & $219.04 \pm 11.18$ & $0.05 \pm 0.00$ & $0.05 \pm 0.00$ \\
& & I.M.(b) & $0.95 \pm 0.00$ & $4539.73 \pm 132.69$ & $228.33 \pm 8.69$ & $191.36 \pm 9.91$ & $0.05 \pm 0.00$ & $0.04 \pm 0.00$ \\
& 2 & G.L.F.(a) & $0.75 \pm 0.01$ & $6265.35 \pm 187.70$ & $620.36 \pm 17.86$ & $553.05 \pm 20.54$ & $0.10 \pm 0.00$ & $0.09 \pm 0.00$ \\
& & G.L.F.(b) & $0.74 \pm 0.01$ & $1743.26 \pm 61.19$ & $592.04 \pm 12.90$ & $518.75 \pm 16.25$ & $0.35 \pm 0.02$ & $0.30 \pm 0.02$ \\
& & I.M.(a) & $0.94 \pm 0.00$ & $7051.31 \pm 190.41$ & $612.50 \pm 23.74$ & $543.62 \pm 27.06$ & $0.09 \pm 0.00$ & $0.08 \pm 0.00$ \\
& & I.M.(b) & $0.94 \pm 0.00$ & $6549.13 \pm 146.01$ & $626.26 \pm 16.04$ & $566.83 \pm 19.62$ & $0.10 \pm 0.00$ & $0.09 \pm 0.00$ \\
& 5 & G.L.F.(a) & $0.74 \pm 0.01$ & $12447.27 \pm 560.25$ & $726.90 \pm 53.55$ & $629.11 \pm 64.83$ & $0.06 \pm 0.01$ & $0.05 \pm 0.01$ \\
& & G.L.F.(b) & $0.74 \pm 0.01$ & $2974.54 \pm 90.05$ & $756.32 \pm 49.99$ & $637.53 \pm 61.54$ & $0.26 \pm 0.02$ & $0.22 \pm 0.02$ \\
& & I.M.(a) & $0.94 \pm 0.00$ & $13331.24 \pm 263.47$ & $1320.95 \pm 77.86$ & $1106.66 \pm 81.97$ & $0.10 \pm 0.01$ & $0.08 \pm 0.01$ \\
& & I.M.(b) & $0.93 \pm 0.00$ & $12501.60 \pm 457.80$ & $1441.09 \pm 54.15$ & $1229.29 \pm 63.45$ & $0.12 \pm 0.01$ & $0.10 \pm 0.01$ \\
\bottomrule
\end{tabular}
    \caption{Comparison of the implicit midpoint and generalized leapfrog integrators on sampling from the posterior of the Fitzhugh-Nagumo model when a randomized number of integration steps are used.}
    \label{tab:randomized-step-fitzhugh-nagumo-ess}
\end{table*}

Results for the Fitzhugh-Nagumo posterior when using the randomized number of steps experimental design are presented in \cref{tab:randomized-step-fitzhugh-nagumo-ess}.

\subsection{Bayesian Logistic Regression}

\begin{table*}[t!]
    \centering
    \scriptsize
    \begin{tabular}{lll|rrrrr}
\toprule
     &      &                            &  Acc. Prob. &   Mean ESS &    Min. ESS &  Mean ESS / Sec. &  Min. ESS / Sec. \\
Step Size & Max. Steps & Method &          &            &            &               &              \\
\midrule
0.1 & 5 & G.L.F.(a) & $1.00 \pm 0.00$ & $286.35 \pm 4.21$ & $208.68 \pm 9.49$ & $3.50 \pm 0.06$ & $2.55 \pm 0.12$ \\
& & G.L.F.(b) & $1.00 \pm 0.00$ & $283.91 \pm 5.17$ & $212.44 \pm 8.27$ & $7.18 \pm 0.16$ & $5.39 \pm 0.25$ \\
& & I.M.(a) & $1.00 \pm 0.00$ & $283.77 \pm 4.26$ & $222.26 \pm 9.88$ & $3.58 \pm 0.06$ & $2.80 \pm 0.13$ \\
& & I.M.(b) & $1.00 \pm 0.00$ & $286.73 \pm 4.68$ & $221.40 \pm 12.74$ & $4.15 \pm 0.07$ & $3.20 \pm 0.19$ \\
& 10 & G.L.F.(a) & $1.00 \pm 0.00$ & $1010.60 \pm 11.69$ & $893.84 \pm 20.10$ & $7.22 \pm 0.10$ & $6.39 \pm 0.16$ \\
& & G.L.F.(b) & $1.00 \pm 0.00$ & $1002.32 \pm 15.93$ & $873.10 \pm 21.47$ & $16.45 \pm 0.27$ & $14.33 \pm 0.35$ \\
& & I.M.(a) & $1.00 \pm 0.00$ & $1005.16 \pm 12.23$ & $871.72 \pm 15.41$ & $7.50 \pm 0.07$ & $6.50 \pm 0.10$ \\
& & I.M.(b) & $1.00 \pm 0.00$ & $1021.59 \pm 8.10$ & $887.65 \pm 27.54$ & $8.48 \pm 0.19$ & $7.37 \pm 0.28$ \\
& 50 & G.L.F.(a) & $1.00 \pm 0.00$ & $14410.27 \pm 147.75$ & $13303.56 \pm 189.57$ & $23.89 \pm 0.19$ & $22.05 \pm 0.26$ \\
& & G.L.F.(b) & $1.00 \pm 0.00$ & $14411.21 \pm 90.58$ & $13450.61 \pm 91.65$ & $60.17 \pm 0.33$ & $56.16 \pm 0.29$ \\
& & I.M.(a) & $1.00 \pm 0.00$ & $14548.15 \pm 121.78$ & $13483.46 \pm 126.40$ & $24.79 \pm 0.19$ & $22.98 \pm 0.22$ \\
& & I.M.(b) & $1.00 \pm 0.00$ & $14519.88 \pm 118.88$ & $13317.24 \pm 127.50$ & $28.52 \pm 0.25$ & $26.16 \pm 0.29$ \\ \midrule
1.0 & 5 & G.L.F.(a) & $0.23 \pm 0.00$ & $904.47 \pm 54.85$ & $542.07 \pm 69.66$ & $4.10 \pm 0.23$ & $2.45 \pm 0.31$ \\
& & G.L.F.(b) & $0.23 \pm 0.00$ & $900.64 \pm 23.10$ & $596.22 \pm 40.82$ & $12.66 \pm 0.37$ & $8.37 \pm 0.56$ \\
& & I.M.(a) & $0.89 \pm 0.00$ & $12294.41 \pm 127.45$ & $10991.38 \pm 217.01$ & $47.59 \pm 0.55$ & $42.55 \pm 0.87$ \\
& & I.M.(b) & $0.89 \pm 0.00$ & $12506.83 \pm 182.42$ & $11280.48 \pm 316.77$ & $50.82 \pm 0.68$ & $45.82 \pm 1.21$ \\
& 10 & G.L.F.(a) & $0.19 \pm 0.00$ & $663.43 \pm 35.45$ & $417.15 \pm 38.40$ & $2.25 \pm 0.11$ & $1.41 \pm 0.12$ \\
& & G.L.F.(b) & $0.19 \pm 0.00$ & $568.08 \pm 31.76$ & $333.88 \pm 39.54$ & $6.32 \pm 0.31$ & $3.69 \pm 0.40$ \\
& & I.M.(a) & $0.87 \pm 0.00$ & $8312.69 \pm 53.82$ & $7449.40 \pm 86.13$ & $17.09 \pm 0.23$ & $15.31 \pm 0.23$ \\
& & I.M.(b) & $0.88 \pm 0.00$ & $8306.90 \pm 46.45$ & $7407.88 \pm 136.86$ & $18.24 \pm 0.19$ & $16.27 \pm 0.36$ \\
& 50 & G.L.F.(a) & $0.10 \pm 0.00$ & $218.25 \pm 33.24$ & $119.54 \pm 22.69$ & $0.41 \pm 0.06$ & $0.22 \pm 0.04$ \\
& & G.L.F.(b) & $0.10 \pm 0.00$ & $199.52 \pm 25.14$ & $110.61 \pm 21.56$ & $1.26 \pm 0.15$ & $0.70 \pm 0.13$ \\
& & I.M.(a) & $0.85 \pm 0.00$ & $6422.28 \pm 65.69$ & $5107.53 \pm 212.97$ & $2.63 \pm 0.04$ & $2.09 \pm 0.08$ \\
& & I.M.(b) & $0.85 \pm 0.00$ & $6693.87 \pm 72.47$ & $5565.30 \pm 272.50$ & $2.90 \pm 0.05$ & $2.41 \pm 0.13$ \\
\bottomrule
\end{tabular}
    \caption{Comparison of the implicit midpoint and generalized leapfrog integrators on sampling from the Bayesian logistic regression posterior on the Breast Cancer dataset when using a randomized number of integration steps.}
    \label{tab:randomized-step-logistic-regression-breast-cancer}
\end{table*}
\begin{table*}[t!]
    \centering
    \scriptsize
    \begin{tabular}{lll|rrrrr}
\toprule
     &      &                            &  Acc. Prob. &   Mean ESS &    Min. ESS &  Mean ESS / Sec. &  Min. ESS / Sec. \\
Step Size & Max. Steps & Method &          &            &            &               &              \\
\midrule
0.1 & 5 & G.L.F.(a) & $1.00 \pm 0.00$ & $286.98 \pm 4.54$ & $234.68 \pm 7.46$ & $2.97 \pm 0.05$ & $2.43 \pm 0.07$ \\
& & G.L.F.(b) & $1.00 \pm 0.00$ & $280.75 \pm 3.93$ & $214.84 \pm 8.31$ & $6.39 \pm 0.09$ & $4.89 \pm 0.19$ \\
& & I.M.(a) & $1.00 \pm 0.00$ & $285.13 \pm 6.19$ & $227.96 \pm 13.22$ & $2.73 \pm 0.07$ & $2.18 \pm 0.13$ \\
& & I.M.(b) & $1.00 \pm 0.00$ & $273.54 \pm 6.51$ & $217.14 \pm 11.43$ & $2.99 \pm 0.07$ & $2.37 \pm 0.12$ \\
& 10 & G.L.F.(a) & $1.00 \pm 0.00$ & $1015.63 \pm 14.83$ & $877.03 \pm 31.16$ & $6.05 \pm 0.09$ & $5.23 \pm 0.19$ \\
& & G.L.F.(b) & $1.00 \pm 0.00$ & $1009.53 \pm 9.46$ & $862.28 \pm 17.84$ & $14.88 \pm 0.13$ & $12.72 \pm 0.31$ \\
& & I.M.(a) & $1.00 \pm 0.00$ & $987.06 \pm 10.31$ & $832.63 \pm 24.35$ & $5.44 \pm 0.07$ & $4.59 \pm 0.13$ \\
& & I.M.(b) & $1.00 \pm 0.00$ & $1010.27 \pm 9.23$ & $867.30 \pm 19.17$ & $6.38 \pm 0.11$ & $5.47 \pm 0.10$ \\
& 50 & G.L.F.(a) & $1.00 \pm 0.00$ & $14735.31 \pm 75.07$ & $13759.10 \pm 141.77$ & $20.13 \pm 0.17$ & $18.80 \pm 0.26$ \\
& & G.L.F.(b) & $1.00 \pm 0.00$ & $14774.00 \pm 99.12$ & $13680.31 \pm 124.54$ & $54.81 \pm 0.38$ & $50.75 \pm 0.49$ \\
& & I.M.(a) & $1.00 \pm 0.00$ & $14748.16 \pm 100.96$ & $13445.33 \pm 111.76$ & $18.49 \pm 0.13$ & $16.85 \pm 0.14$ \\
& & I.M.(b) & $1.00 \pm 0.00$ & $14721.71 \pm 120.84$ & $13834.19 \pm 162.68$ & $21.64 \pm 0.16$ & $20.34 \pm 0.26$ \\ \midrule
1.0 & 5 & G.L.F.(a) & $0.68 \pm 0.00$ & $7161.52 \pm 72.09$ & $6678.78 \pm 98.85$ & $37.13 \pm 0.33$ & $34.63 \pm 0.48$ \\
& & G.L.F.(b) & $0.68 \pm 0.00$ & $7132.61 \pm 58.58$ & $6617.77 \pm 98.49$ & $118.67 \pm 1.61$ & $110.15 \pm 2.30$ \\
& & I.M.(a) & $0.97 \pm 0.00$ & $16750.80 \pm 134.04$ & $15466.43 \pm 76.75$ & $49.44 \pm 0.42$ & $45.65 \pm 0.26$ \\
& & I.M.(b) & $0.97 \pm 0.00$ & $16635.79 \pm 131.54$ & $15402.22 \pm 176.98$ & $51.11 \pm 0.39$ & $47.31 \pm 0.44$ \\
& 10 & G.L.F.(a) & $0.68 \pm 0.00$ & $6498.21 \pm 53.62$ & $5885.25 \pm 55.32$ & $18.86 \pm 0.19$ & $17.08 \pm 0.20$ \\
& & G.L.F.(b) & $0.68 \pm 0.00$ & $6655.24 \pm 61.17$ & $6222.74 \pm 75.56$ & $67.37 \pm 0.79$ & $62.99 \pm 0.91$ \\
& & I.M.(a) & $0.97 \pm 0.00$ & $10858.03 \pm 109.54$ & $10245.67 \pm 118.98$ & $17.69 \pm 0.18$ & $16.70 \pm 0.20$ \\
& & I.M.(b) & $0.97 \pm 0.00$ & $10672.12 \pm 108.02$ & $10006.57 \pm 128.94$ & $18.54 \pm 0.20$ & $17.38 \pm 0.24$ \\
& 50 & G.L.F.(a) & $0.66 \pm 0.00$ & $4999.26 \pm 50.94$ & $4518.13 \pm 54.31$ & $3.41 \pm 0.04$ & $3.08 \pm 0.04$ \\
& & G.L.F.(b) & $0.66 \pm 0.00$ & $5098.68 \pm 72.88$ & $4601.08 \pm 121.45$ & $13.05 \pm 0.27$ & $11.78 \pm 0.36$ \\
& & I.M.(a) & $0.97 \pm 0.00$ & $9176.25 \pm 69.76$ & $8681.47 \pm 129.00$ & $3.27 \pm 0.03$ & $3.10 \pm 0.05$ \\
& & I.M.(b) & $0.97 \pm 0.00$ & $9176.41 \pm 58.95$ & $8657.74 \pm 87.61$ & $3.44 \pm 0.03$ & $3.24 \pm 0.03$ \\
\bottomrule
\end{tabular}
    \caption{Comparison of the implicit midpoint and generalized leapfrog integrators on sampling from the Bayesian logistic regression posterior on the Diabetes dataset when using a randomized number of steps.}
    \label{tab:randomized-step-logistic-regression-diabetes}
\end{table*}
\begin{table*}[t!]
    \centering
    \scriptsize
    \begin{tabular}{lll|rrrrr}
\toprule
     &      &                            &  Acc. Prob. &   Mean ESS &    Min. ESS &  Mean ESS / Sec. &  Min. ESS / Sec. \\
Step Size & Max. Steps & Method &          &            &            &               &              \\
\midrule
0.1 & 5 & G.L.F.(a) & $1.00 \pm 0.00$ & $288.33 \pm 4.45$ & $213.23 \pm 12.18$ & $2.74 \pm 0.05$ & $2.03 \pm 0.12$ \\
& & G.L.F.(b) & $1.00 \pm 0.00$ & $285.05 \pm 2.89$ & $208.93 \pm 12.85$ & $6.13 \pm 0.05$ & $4.49 \pm 0.28$ \\
& & I.M.(a) & $1.00 \pm 0.00$ & $284.53 \pm 2.43$ & $200.77 \pm 9.67$ & $2.86 \pm 0.02$ & $2.02 \pm 0.10$ \\
& & I.M.(b) & $1.00 \pm 0.00$ & $282.89 \pm 4.35$ & $209.47 \pm 7.53$ & $3.26 \pm 0.05$ & $2.41 \pm 0.08$ \\
& 10 & G.L.F.(a) & $1.00 \pm 0.00$ & $1026.88 \pm 6.25$ & $877.98 \pm 13.29$ & $5.50 \pm 0.08$ & $4.70 \pm 0.08$ \\
& & G.L.F.(b) & $1.00 \pm 0.00$ & $1010.32 \pm 7.44$ & $840.99 \pm 18.85$ & $13.45 \pm 0.24$ & $11.20 \pm 0.33$ \\
& & I.M.(a) & $1.00 \pm 0.00$ & $1023.10 \pm 9.64$ & $857.23 \pm 17.40$ & $5.89 \pm 0.05$ & $4.94 \pm 0.09$ \\
& & I.M.(b) & $1.00 \pm 0.00$ & $1026.88 \pm 11.64$ & $865.57 \pm 27.24$ & $6.75 \pm 0.10$ & $5.70 \pm 0.21$ \\
& 50 & G.L.F.(a) & $0.99 \pm 0.00$ & $14285.39 \pm 94.94$ & $13034.74 \pm 212.10$ & $17.93 \pm 0.16$ & $16.36 \pm 0.28$ \\
& & G.L.F.(b) & $1.00 \pm 0.00$ & $14327.81 \pm 86.32$ & $13322.33 \pm 153.96$ & $49.54 \pm 0.37$ & $46.07 \pm 0.61$ \\
& & I.M.(a) & $1.00 \pm 0.00$ & $14465.89 \pm 123.72$ & $13374.24 \pm 126.08$ & $19.04 \pm 0.22$ & $17.60 \pm 0.21$ \\
& & I.M.(b) & $1.00 \pm 0.00$ & $14527.88 \pm 110.84$ & $13225.19 \pm 143.07$ & $22.29 \pm 0.16$ & $20.29 \pm 0.18$ \\ \midrule
1.0 & 5 & G.L.F.(a) & $0.09 \pm 0.00$ & $348.13 \pm 17.73$ & $221.83 \pm 20.56$ & $1.35 \pm 0.08$ & $0.86 \pm 0.08$ \\
& & G.L.F.(b) & $0.09 \pm 0.00$ & $302.71 \pm 17.13$ & $194.10 \pm 18.07$ & $4.44 \pm 0.25$ & $2.84 \pm 0.25$ \\
& & I.M.(a) & $0.76 \pm 0.00$ & $8260.70 \pm 56.57$ & $6971.22 \pm 181.18$ & $21.65 \pm 0.37$ & $18.28 \pm 0.62$ \\
& & I.M.(b) & $0.76 \pm 0.00$ & $8376.43 \pm 83.15$ & $6986.00 \pm 196.62$ & $22.27 \pm 0.33$ & $18.59 \pm 0.61$ \\
& 10 & G.L.F.(a) & $0.06 \pm 0.00$ & $201.29 \pm 13.39$ & $121.49 \pm 12.69$ & $0.69 \pm 0.04$ & $0.42 \pm 0.04$ \\
& & G.L.F.(b) & $0.06 \pm 0.00$ & $212.65 \pm 19.55$ & $107.86 \pm 17.94$ & $2.68 \pm 0.24$ & $1.37 \pm 0.23$ \\
& & I.M.(a) & $0.72 \pm 0.00$ & $5371.69 \pm 83.35$ & $4242.38 \pm 174.21$ & $6.51 \pm 0.12$ & $5.15 \pm 0.23$ \\
& & I.M.(b) & $0.71 \pm 0.00$ & $5464.59 \pm 73.10$ & $4472.08 \pm 119.62$ & $6.82 \pm 0.10$ & $5.58 \pm 0.16$ \\
& 50 & G.L.F.(a) & $0.02 \pm 0.00$ & $54.25 \pm 4.75$ & $16.96 \pm 3.01$ & $0.15 \pm 0.01$ & $0.05 \pm 0.01$ \\
& & G.L.F.(b) & $0.02 \pm 0.00$ & $63.55 \pm 6.00$ & $21.88 \pm 5.09$ & $0.65 \pm 0.06$ & $0.22 \pm 0.05$ \\
& & I.M.(a) & $0.64 \pm 0.00$ & $3384.35 \pm 162.98$ & $2354.23 \pm 228.25$ & $0.57 \pm 0.03$ & $0.40 \pm 0.04$ \\
& & I.M.(b) & $0.63 \pm 0.00$ & $3381.84 \pm 58.00$ & $2217.38 \pm 190.97$ & $0.59 \pm 0.01$ & $0.39 \pm 0.03$ \\
\bottomrule
\end{tabular}
    \caption{Comparison of the implicit midpoint and generalized leapfrog integrators on sampling from the Bayesian logistic regression posterior on the heart disease dataset when using a randomized number of steps.}
    \label{tab:randomized-step-logistic-regression-heart}
\end{table*}

Results for the Bayesian logistic regression models with a randomized number of steps are presented in \cref{tab:randomized-step-logistic-regression-breast-cancer,tab:randomized-step-logistic-regression-diabetes,tab:randomized-step-logistic-regression-heart}.

\end{document}